\pdfoutput=1
\documentclass[%
 reprint,
 amsmath,amssymb,
 aps,
 prx,
]{revtex4-2}

\usepackage{stmaryrd}
\usepackage{graphicx}% Include figure files
\usepackage{dcolumn}% Align table columns on decimal point
\usepackage{bm}% bold math
\usepackage{bbold}
\usepackage{lipsum} 
\usepackage{siunitx}
\usepackage{booktabs}
\usepackage{multirow}
\usepackage{placeins}
\newcommand{\dd}{\mathrm{d}}

\begin{document}

\preprint{}

\title{Physics of anticipatory active matter, \\ with application to crowd dynamics}% Force line breaks with \\

\author{Alexis RAULIN--FOISSAC}
\email{alexis.raulin-foissac@univ-lyon1.fr}
\author{Alexandre NICOLAS}%
\email{alexandre.nicolas@cnrs.fr}
\affiliation{%
  Université Claude Bernard Lyon 1, Institut Lumière Matière, CNRS, UMR 5306, 69100, Villeurbanne, France
}%

\newcommand{\stateA}{\widetilde{\boldsymbol{Q}}}
\newcommand{\trajA}{\widetilde{\boldsymbol{q}}}
\newcommand{\costA}{\widetilde{\mathcal{C}}^{\omega}}
\newcommand{\state}{\boldsymbol{q}}
\newcommand{\past}{\boldsymbol{Q}}
\newcommand{\NN}{\mathcal{N}}
\newcommand{\PeBare}{\textit{Pe}}
\newcommand{\PeDressed}{\mathrm{Pe}}
\newcommand{\TantBare}{T_{ant}}
\newcommand{\TantEff}{\mathrm{T}_{\mathrm{ant}}}

\date{\today}% It is always \today, today,

\begin{abstract}
\begin{description}
\item[Context]
Statistical Physics has traditionally dealt with entities that interact merely based on the present, and possibly past, configurations. This reactive framework is inefficient in many situations involving living beings, such as predators chasing a prey, pedestrians, or even robots. 

\item[Results] This paper introduces a statistical physical framework for the dynamics of anticipatory agents, whose present-time dynamics depend on the prospective system state that they anticipate.
We clarify how these dynamics can be expressed in terms of a cost function constructed based on observations and we show that the dynamics of an anticipatory agent in $d$ dimensions can be mapped onto the dynamics of a (non-anticipatory) chain in $d+1$ dimensions, with fluctuations acting transversely on the chain to account for the uncertainty about the future state. Insights from polymer Physics help us characterize the dynamics of these chains and delineate an anticipation horizon beyond which the blurry future can be handled in a mean-field way.

\item[Application]
The foregoing framework is successfully applied to pedestrian dynamics, leading to a seamless integration of operational and tactical levels  in an agent-based model.
Even with a minimal expression of the cost, the model succeeds in reproducing various experimental scenarios which are challenging for state-of-the-art models, such as crossing cluttered environments or alighting from a crowded train. The transparent and flexible basis of the model allows the straightforward incorporation of additional mechanisms.

\end{description}
\end{abstract}

\maketitle

%\tableofcontents
\section{\label{sec:intro}Introduction}
Anticipation refers to ``actions taken in relation to an event that has not occurred yet'' \cite{adrian2025glossary}. 
A putative cornerstone in the singular evolution of the hominids \cite{bownds1999biology}, the development of anticipatory capabilities has also benefited multiple other natural species, from large predators and bees down to bacteria \cite{suddendorf2022anticipation}, as well as robotic systems \cite{van2008reciprocal,ratliff2009chomp}. 

Since the end of the 20th century, Statistical Physics has triumphally extended its realm to the dynamics of living systems and active matter \cite{marchetti2013hydrodynamics}, but it has never fully severed the bonds with its reactive (non-anticipative) roots, perhaps saddled by an all-too-narrow vision of causality. In particular, while the effect of perception and the ensuing non-reciprocity of interactions between agents have progressively come in the limelight \cite{lavergne2019group,ivlev2015statistical}, it is usually taken for granted that these interactions should be formulated based on only the present configuration of the system (and past ones, if the system has memory).

The insufficiency of this standpoint for practical purposes has reached a breaking point in recent years, notably (but not only) in the field of pedestrian dynamics. Empirical crowd datasets have shown that what governs inter-pedestrian interactions is hardly relative positions only, but rather anticipated times to collision (TTC), namely, the delay after which a collision is to be anticipated if all velocities are conserved \cite{Karamouzas2014universal,cordes2023dimensionless}. The response of a static crowd to an intruder's crossing also largely hinges on anticipation \cite{bonnemain2023pedestrians}.

To remedy the problems of reactive models, a number of models for pedestrian dynamics have introduced a modicum of anticipation, by adding \emph{ad hoc} terms that mimic specific anticipatory responses \cite{xu2021anticipation,wolinski2016warpdriver}, or by making interactions TTC-dependent \cite{Karamouzas2014universal,echeverria2023body} or (in a similar vein) by barring the selection of velocities that would lead to a collision \cite{van2008reciprocal,van2011reciprocal}. The latter options have succeeded in better reproducing salient features of crowd dynamics, but nevertheless they rely on a linear extrapolation of trajectories, independent of one's own actions, and thus do not strongly depart from the classical reactive theoretical framework; other examples of low-order accounts of anticipatory responses, in other self-propelled assemblies, can be found in \cite{morin2015collective,szabo2009transitions}.
% second ref can be removed if needed
For several practical situations, e.g., crossing a cluttered space, these extensions were shown to be insufficient \cite{raulin2025highs}.

Poles apart from these linear extrapolations, optimal control and game theory aim to capture full trajectories that are optimal in some sense. They have been applied to crowd dynamics with notable success \cite{hoogendoorn2003simulation,ziebart2009planning,zhi2021anticipatory,modi2023mutiagent}, but for large anticipatory assemblies they give rise to such technical complications that the ensuing collective motion  eludes any intuitive grasp. Mean-field approaches overcome these technical issues to some extent, but at the substantial expense of obliterating the agents' granularity, heterogeneity and variability, and assuming omniscient pedestrians.
Further issues arise from more technical considerations, such as putting on an equal footing multiple local solutions (Nash equilibria) and looking for optimal trajectories that extend over arbitrarily long time.

In this contribution, we wish to establish a statistical physical framework for the dynamics of anticipatory agents that is not marred by the opacity of current game-theoretical approaches. We also aspire to anchor this approach in a firmer basis, where (instead of positing utilities or costs) the equations are framed based on observations and agents are not supposed to be rational from the outset.
These goals are not achieved by perturbatively amending the classical reactive framework, but by including long-term anticipation from scratch. Indeed, we believe that the strategy of jumping in the deep end may be more successful than making small perturbative steps: for instance, Statistical Mechanics can hardly be derived by moving from one deterministic trajectory to a few interacting trajectories, but instead by considering infinitely many. 

Here, after formalizing the notion of cost as an observation-based construct, we delineate the dynamics of reactive agents \emph{vs.} anticipatory ones, whose present-time motion depends on the prospective state of the system that they anticipate, in Sec.~\ref{sec:Formalization_anticipation}. We show that anticipatory dynamics in $d$ dimensions can be mapped onto the dynamics of a (non-anticipatory) chain in $d+1$ dimensions, with fluctuations acting transversely on the chain to account for the uncertainty about the future state.
The implications of this framework are made more tangible in Sec.~\ref{sec:level3} by considering the dynamics of a single goal-driven agent; this leads to the definition of an anticipation horizon, beyond which the blurry future is handled in a mean-field way and before which trajectories are discretized in time, thus turning into polymer-like chains.
In Sec.~\ref{sec:Insights_CondMat}, insights from polymer Physics help us characterize the dynamics of these chains and also guide the numerical implementation of the model; note that this section is mostly technical and can be skipped for cursory reading.
Finally, Sec.~\ref{sec:level4} applies the proposed general framework to pedestrian crowds and demonstrates that, even with a straightforward implementation and simple cost functions, the resulting agent-based model is able to replicate empirical scenarios with crowds that are particularly challenging for state-of-the-art models. Readers only interested in this disciplinary application can directly jump to this section.

\section{\label{sec:Formalization_anticipation}Formal definition and theory of anticipatory dynamics}

Our ambition is to develop a dynamical theory of active systems that integrates anticipation, in an intuitive framework that makes efficient numerical simulations possible. This ambition manifestly requires a clear delineation of anticipatory systems.

\subsection{Dynamics of interacting agents}
Interacting agents such as animals or human pedestrians do not traditionally pertain to Statistical Physics, so we start by contextualizing their dynamics on the basis of objective, observable quantities. 
For this purpose, let us consider an assembly of $\NN$ agents indexed by $j=1\dots  \NN$, whose trajectories are globally denoted by $\past^{\omega}=\Big({\state}_{j}^{\omega}\Big )_{j=1\dots \NN}$, with ${\state}_{j}^{\omega}(t)$ representing the position of agent $j$ in $d$ dimensions at time $t\in\,[0,T]$. These trajectories may vary between realizations $\omega$, hence the superscript. For each realization $\omega$, it is always possible to design a function $\mathcal{C}^{\omega,j}$, such that 
\begin{align}
    {\past}^{\omega} = \underset{\past\in\mathbb{R}^{d\NN\times [0,T]}}{\arg\min}~ \mathcal{C}^{\omega}[\past],
    \label{eq:argmin_cost0}
\end{align}
where adequate initial conditions are prescribed. The goal of a dynamical description is to capture the statistics of $\past^{\omega}$ over realizations with an \emph{explicit} function $\mathcal{C}[\past]$ (instead of $\mathcal{C}^{\omega}[\past]$), possibly complemented with an \emph{explicit} random distribution to account for the variations between realizations. For example, for a Newtonian particle of mass $m$ in a potential $U$, one can make use of 
\begin{equation}
    \mathcal{C}[\past]=\int_0^T \sum_j \Big(m\ddot{\state}_{j}(t) + \frac{\partial U}{\partial  {\state}_{j}}\Big)^2 dt \nonumber
\end{equation}
or substitute it with the usual  action $\int_0^T \sum_j \Big(\frac{1}{2}m\dot\state_j^2(t)-U\big(\past (t)\big)\Big)dt$ if the end positions are fixed, whereas overdamped particles exposed to a friction coefficient $\zeta$ are suitably described by
\begin{equation}
\mathcal{C}[\past]=\int_0^T \sum_j \Big(\zeta\dot{\state}_{j}(t) +  \frac{dU}{d {\state}_{j}}\Big)^2 dt. \nonumber
\end{equation}

The use of a global function $\mathcal{C}[\past]$ entails a reciprocity principle ($\frac{\partial }{\partial {\state}_{j}} \frac{\partial  \mathcal{C}}{\partial {\state}_{i}}=\frac{\partial }{\partial {\state}_{i}} \frac{\partial  \mathcal{C}}{\partial {\state_j}}$, for smooth enough functions) that holds for simple particles, but not for general agents, owing \emph{inter alia} to the asymmetry of perception. This issue is overcome by looking for explicit \emph{agent-dependent} functions $\mathcal{C}_j[\past]$ (in lieu of $\mathcal{C}[\past]$), possibly including random terms,
whose ``minimizers'' $\past^{\star}$, defined by the Nash equilibrium conditions
\begin{align}
    \forall j,\  {\state}_{j}^{\star}= \underset{ \state_j \in\mathbb{R}^{d\times [0,T]} }{\arg\min}~ \mathcal{C}_j[{\state}_{1}^{\star},\dots,{\state}_{j},\dots,{\state}_{\NN}^{\star}],
    \label{eq:argmin_cost1}
\end{align}
are statistically similar to $\past^{\omega}$.

\subsection{Anticipatory systems}
To clarify the notion of anticipation, consider an arbitrary intermediate time $t$.
For \emph{reactive} systems, the trajectories up to $t$, ${\past}^{\omega}_{t''\leqslant t}$, can be
determined independently of the motion planned for $t'>t$. 
Accordingly, in Eq.~\ref{eq:argmin_cost1}, it is sufficient to look for functions $\mathcal{C}_{j}$ that depend on test trajectories $\past_{t''\leqslant t}$ extending only up to $t$. In particular, the dynamics at time $t$ are ruled by quantities measured up to time $t$ (or just at time $t$ for systems without memory), viz.,
\begin{equation}
  {\textsc{\footnotesize [Reactive agents]}}\ \ \forall j,\  \frac{\delta \mathcal{C}_j}{\delta {\state}_{j}(t)}\Big[\past_{t''\leqslant t}={\past}^{\omega}_{t''\leqslant t}\Big] = 0.
  \label{eq:reactive}
\end{equation}
For the aforementioned overdamped particle, it is easy to check that Eq.~\ref{eq:reactive} boils down to the overdamped equation
\begin{equation}
\forall j,\  \zeta \dot{\state}^{\omega}_{j}(t) = - \frac{\partial U}{\partial {\state}_{j}}({\past}^{\omega}_{t}). \nonumber
  \label{eq:reactive_overdamped}
\end{equation}
Beyond this specific example, simple models of self-propelled particles, such as active Brownian particles, also naturally fit in this reactive category.

By contrast, the dynamics of \emph{anticipating} agents seem to be governed by how they predict the system will evolve: a predator chasing a prey is not interested in where the prey currently is, but in where it will soon be; many social bees start to become more active just before (i.e., in anticipation of) sunrise, while still inside the nest cavity \cite{bloch2017time}; a pedestrian does not fear collisions with pedestrians at their \emph{current} positions, but where they should be next \cite{Karamouzas2014universal,meerhoff2018guided}; a person at a bus stop does not care about where the bus currently is, but where and when it will stop; even \emph{E. Coli} bacteria  traveling through human digestive tracts activate genes for maltose digestion not necessarily because they currently are in maltose-rich regions, but before reaching (i.e., in anticipation of) such regions \cite{mitchell2009adaptive,suddendorf2022anticipation}. 
More generally, Robert Rosen \cite{rosen2011anticipatory} claims that biology in general cannot be reduced to just applied physics, because living organisms notably have a distinctive capacity to anticipate: they do not merely react to the present but adjust their behavior to the (broadly defined) anticipated state $\stateA^{\omega, j}(t')$ that they predict for future times $t'$ in light of the current and past states $\past^{\omega}_{t''\leqslant t}$ e.g. via an internal model of themselves and their environment.

Accordingly, the reactive framework of Eq.~\ref{eq:reactive} is inefficient for anticipating agents.
%A system is thus termed anticipatory if it contains a model of itself and/or its environment, and if this model can influence its present behavior so as to account for future states.
On the other hand, at time $t$, agents $j$ do not have access to the exact future trajectories ${\state}_{i}^{\star}$ of other agents $i$ featured in Eq.~\ref{eq:argmin_cost1}. Therefore, we substitute the future trajectories on the rhs of Eq.~\ref{eq:argmin_cost1} with the anticipated state $\stateA^{\omega, j}$, arriving at 
\begin{align}
 {\textsc{\footnotesize [Anticipatory agents]}} \nonumber &\\
    \forall j,\ {\state}_{j}^{\star}= \underset{ \state_j \in \mathbb{R}^{d\times [0,T]}}{\arg\min}~ &\mathcal{C}_j[{\state}_{j},{\past}^{\star}_{t''\leqslant t},\stateA^{\omega, j}({\past}^{\star}_{t''\leqslant t},\state_j)],
    \label{eq:argmin_cost2}
\end{align}
where we have made explicit the dependence of the anticipated state on the surroundings' state ${\past}^{\star}_{t''\leqslant t}$ and on the prospective trajectory $\state_j$ under consideration.
In particular, at time $t$,
\begin{align}
  {\textsc{\footnotesize [Anticipatory agents]}}& \nonumber\\ \ \ \forall j,\  \frac{\delta \mathcal{C}_j}{\delta {\state}_{j}(t)} & [\state_j,{\past}^{\omega}_{t''\leqslant t},\stateA^{\omega, j}]\Big|_{\state_j=\state_j^{\star}} = 0.
 \label{eq:vitesse}
\end{align}
(In practice, however, $\state_j^{\star}$ is obtained from Eq.~\ref{eq:argmin_cost2}.)

  The internal representation $\stateA^{\omega, j}$ is not directly observable; yet, it is a convenient proxy for the variables that guide decision-making, somewhat similarly to the use of complex numbers to encode phase and dynamical information in Electromagnetism, although physical observables are all real numbers. What is observable here is the agent's actions $\state_j^{\star}(t+\epsilon)$ when $\epsilon \to 0^{+}$, i.e., how they ``step into the future''. Unless something unexpected happens, this action should be consistent with the trajectory that the agent has anticipated for themselves in the imminent future, i.e., $\trajA^{\omega, j}_j (t+\epsilon)= \state_j^{\star}(t+\epsilon)=\state_j^{\omega}(t+\epsilon)$.

\subsection{From a multi-verse of anticipated states to a shared anticipation universe}
\label{sec:shared-space}

Overall, solving the assembly's dynamics via Eq.~\ref{eq:argmin_cost2} requires continuously updating a multi-verse, or more precisely an $\NN$-verse, of anticipated states $\Big\{{\stateA}^{\omega,j}\Big\}_{j=1\dots \NN}$. Indeed, different agents need not have identical anticipations. To better grasp the complexity of the problem, a loose parallel may be drawn with the paradox of Wigner's friend in Quantum Mechanics
\footnote{The paradox of Wigner's friend refers to a situation in which ``Wigner'' observes a friend of his perform a measurement on a quantum object. Depending on the viewpoint, the wavefunction of the object was reduced either during the friend's measurement or when Wigner learned about the result, which is paradoxical if this reduction is an objective event.}, if the prediction ${\stateA}^{\omega,j}$ is probabilistic (i.e., a collection of density functions, rather than trajectories): pending a future observation, agent $j$ (like Wigner) is uncertain about their `friend' $i$'s position (even though $i$ knows it) and must \emph{blindly} evolve their own prediction, based on their internal model, until an observation determines agent $i$'s actual state. The resulting assembly of $\NN$ Wigner's friends, who mutually observe each other every now and then and update their own anticipation universe ${\stateA}^{\omega,j}$ accordingly, is practically intractable for $\NN\gg 1$.

However, if the observations are frequent (which will be the case for pedestrians), it makes sense to consider that the present-time state $\stateA^{\omega,j}(t'=t)$ is the configuration of all the agents that are currently perceived by $j$. Furthermore, anticipation will only be efficient if, for all perceived agents $i$, agent $j$ predicts trajectories $\trajA_i^{\omega,j}(t')$ that closely match the actual ones in the very near future $t' \approx t$,  
$\trajA_i^{\omega,j}(t') \approx \state^\omega_i(t') \approx \trajA_i^{\omega,i}(t')$. This implies that the internal models of agents $j$ yield similar short-term predictions for the agents that are perceived. Therefore, we recast the state anticipated by agent $j$ into
\begin{align}
\stateA^{\omega,j} (t')= \Big\{\pi^j_{t'}\left(\trajA^{\omega}_i\right)\Big\}_{i=1\dots \NN}.
\label{eq:perception_mapping}
\end{align}
Here, the observer-specific functions $\pi^j_{t'}$  \emph{warp} \cite{wolinski2016warpdriver} the self-centered predictions  $\trajA^{\omega}_i\hat{=}\trajA^{\omega,i}_i$, in order to account for nonperception ($\pi^j_{t'}\left(\trajA^{\omega}_i\right)=\varnothing$) and for the probable divergence of predictions at long times.
(In practice, we will use extremely simple warping functions $\pi^j$). 

The simplifying assumptions leading to Eq.~\ref{eq:perception_mapping} have a major practical advantage: they reduce the anticipation multi-verse    $\Big\{{\stateA}^{\omega,j}\Big\}$ to a shared base universe 
$\stateA^{\omega}\hat{=}\Big\{\trajA^{\omega}_i\Big\}_{i=1\dots \NN}$, possibly warped by the observer's perspective and internal rules. While  $\Big\{{\stateA}^{\omega,j}\Big\}$ consists of $\NN$ states of multiple (up to $\NN$) predicted trajectories each, the shared base universe $\stateA^{\omega}$ consists of only $\NN$ trajectories.  Naturally, this comes at a cost: observer-dependent correlations, whereby for instance $j$ mispredicts $i$'s trajectory because $i$ actually interacts with another agent not seen by $j$ (or the other way round), are barred. That being said, even in the shared universe, the  trajectory that agent $j$ predicts for $i$ does not depend solely on the past, but possibly also on the future actions contemplated by $j$ -- a crucial point for what follows.

\subsection{Variability between realizations}
As for the actual trajectories, for each realization $\omega$, it is always possible to design functions $\costA_j$ that are minimized by the planned trajectories $\trajA^\omega_j$,  

\begin{align}
     \forall j,\ \trajA^\omega_j = \underset{\state_j}{\arg\min}~ \costA_j\Big[\state_j |\left\{\pi^j(\trajA^{\omega}_i)\right\}_{i\neq j}\Big].
    \label{eq:argmin_cost}
\end{align}
$\costA_j$ can be regarded as the `cost' that agent j aims to minimize and actually minimizes, for a given set of others' anticipated trajectories $\left\{\pi^j(\trajA_i)\right\}_{i\neq j}$. 
%Note that since others agents do the same, this is indeed a game theoretical setup. 
The costs $\costA_j$ vary between realizations $\omega$. We express these variations as (supposedly modest) fluctuations around a central cost function $\mathcal{C}_j[\state_j\,\,|\left\{\pi^j(\state_i)\right\}_{i\neq j}]$, viz., 
\begin{equation}
     \costA_j = \mathcal{C}_j + \epsilon^{\omega}_j.
\end{equation}

Roughly speaking, the fluctuating cost has an effect similar to temperature in a physical system, populating trajectories whose ``costs" are not strict minima of $ \mathcal{C}_j$. Its origin can be ascribed to the agents' variable responses  and their uncertainty about the future; its magnitude should increase with time: the further in time, the more uncertain (until trajectories become fully blurred, and equiprobable, in the  distant future.)
%, meaning that no further insights can be drawn).

Incidentally, the introduction of anticipation in Eq.~\ref{eq:argmin_cost2} and, above all, of a ``cost'' in Eq.~\ref{eq:argmin_cost} puts in the limelight the much-debated issue of the rationality of the agents. One should bear in mind that our reasoning is premised on no such rational hypothesis.  Interpreting the statistics-based (or probability-based) auxiliary function $\mathcal{C}_{j}$ as a \emph{cost}  is a subjective choice here, made out of intellectual convenience. So is the choice to derive the \emph{observed} dynamics using virtual anticipated trajectories and/or a putative internal model of the environment maintained by agents in Eq.~\ref{eq:argmin_cost2}. In Kantian parlance, these choices echo the subjective purposiveness that the Subject introduces in contingent natural forms such as sea shells to make up for the inability of Reason to grasp them, where the ``laws of causality arising from the simple mechanisms of Nature do not suffice for their understanding'' \cite{kant2000critique}.

\subsection{Geometric interpretation of anticipatory dynamics}
\label{sec:IID}
Provided the cost functions are smooth enough, the minimization of Eq.~\ref{eq:argmin_cost} can be achieved by letting the planned trajectory $\trajA_j$ descend the gradient of $ \costA_j$ in artificial time $\tau$, for all $j$:
%planned trajectories $\trajA_j$ can be sampled around local minima of the central costs by a gradient-descent algorithm for all $j$:

\begin{align}
\frac{\partial \trajA_j(t')}{\partial\tau}& =  -\frac{\delta \costA_j[\trajA_j|\pi^j(\trajA_{i\neq j})]}{\delta \trajA_j (t')} \nonumber \\
 & = -\frac{\delta \mathcal{C}_{j}[\trajA_j|\pi^j(\trajA_{i\neq j})]}{\delta \trajA_j (t')} + \eta_j^{\omega},
 \label{eq:Langevin1}
\end{align}
where we have introduced the shorthand $\eta_j^{\omega}=-\frac{\delta \epsilon^{\omega, j}}{\delta \trajA_j}$ and the dependence of $\trajA_j(t')$ on $\tau$ is implicit.
% FOrmerly: $\eta_j^{(\epsilon)}=-\frac{\delta \epsilon^{\omega, j}[\trajA_j|\pi_j(\trajA_{i\neq j})]}{\delta \trajA_j}$
Now, should one interpret $t'$ in Eq.~\ref{eq:Langevin1} as just another dimension (on top of the $d$ spatial dimensions of $\trajA_j$), one will immediately see a parallel between the planned trajectories $\trajA_j$ and chains that wiggle in space-time (see Fig.~\ref{fig:antipodal}a). This is the first take-home message of our paper: the dynamics of anticipatory agents in $d$ dimensions reduce to the dynamics of interacting (but non-anticipatory) chains in $d+1$ dimensions. This formal mapping is reminiscent of the use of Wick's rotation of time to connect Quantum Field Theory to Statistical Field Theory.

\begin{figure}
    \centering
    \includegraphics[width=0.7\linewidth]{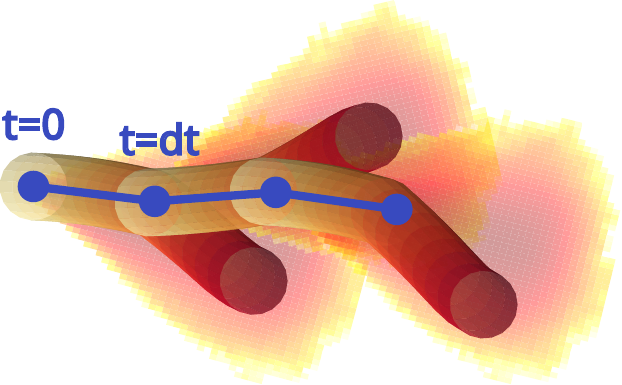}
    
    \caption{Interplay between real-time trajectories (\emph{in blue}) and artificial dynamics (\emph{heatmap}). The thick chains in warm colors (from yellow -- near future, with limited fluctuations -- to dark red -- longer-term future, with higher fluctuations) represent the last chain configuration at the end of the artificial (Langevin) dynamics at successive real-time steps $t=0,\dd t,\dots$ and the heatmaps in the background are the corresponding probability density functions over the whole dynamics. 
    At each real time step, the first position of the chain is frozen and real time is shifted ($t \to t + \dd t$).}
    \label{fig:tubes}
\end{figure}

\subsection{Real-time motion and artificial dynamics}
\label{sec:artificial_dynamics}
To sample planned trajectories, minimizers of $\costA_j$ could in principle be obtained by simulating Eq.~\ref{eq:Langevin1} for a given $\omega$ (hence, $\eta_j^{\omega})$,  collecting the resulting $\trajA_j$ after a long relaxation time $\tau\to\infty$,  running the simulation afresh for another $\omega$, and so on.  However, as the random perturbation $\eta_j^{\omega}$ is unknown, we will overlook this quenched disorder and simply assume that $\eta_j^{\omega}$ is fully uncorrelated in fictitious time $\tau$ and in space, so that it acts as white noise. Equation~\ref{eq:Langevin1} is then akin to an overdamped stochastic Langevin equation. Its numerical resolution enables one to sample the $\trajA_j$'s over artificial time $\tau$, after an ``equilibration'' period (the duration of which will be derived using insights from Polymer Physics in Sec.~\ref{sec:condensed-matter}).

This ``equilibration'' does not produce thermal equilibrium, for two reasons. Firstly, the `forces' acting on a chain do not derive from a global potential and will generally be non-reciprocal. 
%(In passing, note that a dual probabilistic approach may also have been followed to arrive at a similar equation, as in the Langevin Monte Carlo framework, but possibly with a more opaque derivation.)
Secondly, the volatility of $\eta_j^{\omega}$ grows with $t'$, mirroring the increasing uncertainty in the \emph{future} direction: the chain  wiggles in a temperature gradient directed towards the future. 
At the \emph{present} end ($t'=t$), shown in yellow in Fig.~\ref{fig:tubes}, the chain is pinned (its position is fully determined). As time $t$ goes on, the temperature gradient is shifted towards the future and the \emph{present} end of the chain gets frozen at its current position (marked as a blue dot in Fig.~\ref{fig:tubes}). The simulated dynamics thus alternate between `equilibrations' in artificial time $\tau$ (Eq.~\ref{eq:Langevin1}), during which the chain randomly explores  configurations, and stepwise shifts in real time ($t\to t+\delta t$). One can interpret these steps as decision-making processes through which anticipating agents continuously update their planned trajectories.

\subsection{Connections with game theory and motion planning}
\label{sub:connections}
At this stage, some connections ought to be drawn between the newly introduced framework and adjacent fields.

First consider the limit-case of vanishing variability between realizations ($\epsilon^{\omega,\,j} \to 0$) and perfect inference of the others' trajectories (i.e., $\pi^j = \mathbb{1}$ is the identity function or, equivalently, $\stateA^{\omega,j}=\stateA^{\omega}$). In this case, the stationary solutions of the gradient descent Eq.~\ref{eq:Langevin1} satisfy the optimality condition of Eq.~\ref{eq:argmin_cost1}: they are Nash equilibria of a game between $\NN$ players aiming to find a strategy (i.e., a trajectory) that minimizes their cost $\mathcal{C}_j$. 
Thus, although the proposed theory of anticipative dynamics was not originally framed as a  competition between rational agents, it does boil down to a game \emph{in the aforementioned limit}.
More precisely, the shared universe hypothesis   (Sec.~\ref{sec:shared-space}) with perfect inference ($\pi_j = \mathbb{1}$) implies the  common knowledge hypothesis of game theory
\footnote{The converse is not true, as one can readily see with the following counterexample. Suppose that agent $j$ has the choice between two equivalent optimal strategies. The shared universe hypothesis allows other agents to infer agent $j$'s path, whereas in the common knowledge hypothesis other agents only know that agent $j$ will choose either of the strategies.},
as the optimal choice of each agent is known to, and used by, everyone else. 

However, the proposed theory eschews the conundrum of (perfect) rationality via the fluctuating term in Eq.~\ref{eq:Langevin1}. Not only does the notion of a cost emerge as a statistical construct, rather than a postulate on the agents' behaviors, but in addition the planned trajectories are not strict minima of this cost. They are `satisficing' solutions (using the expression coined by Herbert Simon), rather than strict optima, and are thus easier to reach by learning \cite{garnier2024unlearnable}. 

Besides, the random fluctuations also serve a more practical purpose, in that they prevent the indefinite entrapment of trajectories in a given local minimum, making a broader exploration of phase space possible.
In this regard, combining fluctuations with a gradient term in Eq.~\ref{eq:Langevin1} is reminiscent of the inclusion of stochasticity in optimization-based motion planners in the field of robotics. These planners aim to find trajectories in a given workspace for body points on a robot such that a global cost penalizing collisions and jerky motion is minimized. In particular, the CHOMP algorithm follows from writing a gradient descent scheme in covariant form; it complements this gradient descent with random `pushes' generated by a Hamiltonian Monte Carlo algorithm with resetting, in order to push the system out of basins of attraction in which it may be trapped \cite{ratliff2009chomp}. In the alternative STOMP algorithm, stochasticity instead emerges from the computation of gradients based on a sample of directions \cite{ratliff2009chomp}, instead of exact gradients. A connection
\footnote{Interestingly, in \cite{ziebart2009planning}, the authors reverse-engineered related motion-planning ideas to predict probable pedestrian motion in the context of robotic navigation; this was achieved by reinforcement learning, with a model specifically trained for a given environment. But interactions between the anticipatory agents were largely overlooked.}
can thus be drawn between motion planning in robotics and the dynamics of anticipatory agents, even though in the former stochasticity does not reflect uncertainty or limited anticipation capabilities, but only has a practical purpose. Moreover,  a global potential (or cost function) is used in robotics, whereas this is not legitimate for an assembly of autonomous agents with possibly non-reciprocal interactions. 

 With autonomous agents, the game-theoretical optimization process is typically exposed to a curse of dimensionality (i.e., a computational complexity that increases exponentially with the number $\NN$ of agents, insofar as an optimum should be sought in a space of dimension proportional to $\NN$).
 We will see that this problem will be bypassed by simulating Eq.~\ref{eq:Langevin1} in a discretized form, with simple enough (but practically relevant) cost functions, and using insights from Condensed Matter Physics.

\section{\label{sec:level3}Dynamics of a single purposeful agent}
% NB: in the motion-planning literature, the wording is "purposeful" instead of "goal-driven"
\subsection{\label{sec:costDefinition}Generic expression of the cost}
Given the novelty of the proposed framework, we start by analyzing the response of a single purposeful (anticipatory) agent, with a minimal form for the cost functional $\mathcal{C}$. The agent heads for a target area $\state_g$, so there is a penalty $c_g(\widetilde{q})$ for being away from $\state_g$. However, the agent is aware that it cannot move infinitely fast: there must be a cost $c_v(v)$ that becomes prohibitive at large speeds $v \hat{=} ||\dot{\trajA}||$. Thus, upon integration along the agent's planned trajectory,

\begin{align}
    \mathcal{C}[\trajA] = \int_t^{T}\Big[c_v\left(||\dot{\trajA}(t')||\right) + c_g\left(\trajA(t')\right)\Big]\dd t'.
    \label{eq:integral_cost}
\end{align}
The results presented in this section are valid for fairly general functions $c_v$ and $c_g$, as long as $c_v(v)$ increases superinearly with $v$ and that $c_g$ is peaked around $\state_g$. These conditions ensure that, in the competition between the driving term $c_g$ and $c_v$, an optimal trade-off is reached at a \emph{finite} speed $v_{\mathrm{opt}}$. 

In particular, supposing $c_v(v) = \alpha v^\beta$ and $c_g(\trajA)= \kappa(1- \chi_{\state_g}(\trajA))$, where $\alpha,\,\kappa>0$, $\beta>1$ and $\chi_{\state_g}$ is the indicator function of a small neighborhood of $\state_g$, one arrives at 
\begin{align}
    v_{\mathrm{opt}} = \left(\frac{\kappa}{\alpha(\beta-1)}\right)^{1/\beta}.
    \label{eq:v_opt}
\end{align}

%The process of finding the trajectory planned by one agent becomes analogous to the evolution of a polymer subjected to thermal fluctuations, the uncertainties, cohesive forces, bio-mechanical constraints, and attracted to the target.

%Explanation about the difference times, real one and artificial one upon which polymers are equilibrated and how to advance in the real time -> leads to the question of how long do polymers need to equilibrate.  
\subsection{Fluctuations and anticipation horizon}
\label{sec:IIB}
We recall that in the Langevin-like Eq.~\ref{eq:Langevin1}, we model the random contribution $\eta(\tau, t')=\frac{\delta \epsilon^{\omega}}{\delta \trajA}$ as white noise, with an amplitude that grows with $t'$, mirroring a growing uncertainty. Thus, for the single agent, even without any amplification of the uncertainty due to the uncertain behavior of neighbors, the variance over $\tau$ of the planned position $\trajA(\tau,t')$ increases with $t'$, not to mention that the chain is fixed at one end ($t'=t$).

It follows that the chain fluctuations will exceed any typical length of the system (e.g., the agent's size)  past some time $t' = t + \TantBare$; we will refer to $\TantBare$ as the anticipation horizon and evaluate it analytically in Sec.~ \ref{sec:condensed-matter}, after specifying the temperature gradient. Before $\TantBare$, trajectories can be predicted with meaningful precision; the chain is localized. Beyond this horizon, uncertainty dominates.  
%in the case of fluctuations whose amplitude scale quadratically with $(t'-t)$, $< \eta(\tau, t')\eta(\tau', t') >= \Theta (t'-t)^2\delta(\tau - \tau')\delta(t'-t')$.

\subsection{Beyond the anticipation horizon}
\label{sec:beyond-anticipation}
So far, chains (i.e., planned trajectories) can be almost infinitely long, which is impractical and often meaningless. Therefore, we want to approximate the contribution to the cost beyond $\TantBare$ ($\TantBare \ll T$) with an effective `final' cost $\mathcal{C}_f \!\left(\trajA(\TantBare)\right)$ (averaged over fluctuations) that only depends on the agent's position $\trajA(\TantBare)$ on the verge of the anticipation horizon. 
$\mathcal{C}_f$ pulls the chain end toward the target. 
With the expressions of Eq.~\ref{eq:v_opt}, noticing that the instantaneous costs vanish once the agent has reached the target $\state_g$,
\begin{align}
    \mathcal{C}_f\!\left(\trajA\big(\TantBare\big)\right)
    &= \int_{\TantBare}^{T}
    \Big(c_v( v(t')) + c_g(\trajA(t'))\Big)\,\mathrm{d}t' \nonumber\\
    &= \int_{\trajA(\TantBare)}^{\state_g}
    \Big(\alpha  v_{\mathrm{opt}}^\beta + \kappa\Big)
    \frac{\mathrm{d}\ell}{ v_{\mathrm{opt}}} \nonumber \\
    &=  \kappa\frac{\beta}{\beta-1} \frac{D_g}{ v_{\mathrm{opt}}},
\end{align}
% Alexis: j'ai enlevé les cdot car "Do not use a center dot to indicate multiplication except for products of vectors, dyadics, and the like." d'après les guidelines
Thus, in this minimal setting, the final cost is proportional to the effective distance  $D_g$ that remains from $\trajA\big(\TantBare\big)$ to $\state_g$, taking into account the geometry of the premises.

More generally, the local cost functional will include contributions due to anticipated interactions with other agents and a mean-field approximation will be required. Concretely, the individual trajectories $\trajA_j$ will be averaged over fluctuations and replaced by densities of probability, thereby capturing their average influence on a representative agent, but neglecting higher-order correlations.

%At long times, fluctuations become too strong, so the infinite polymer is split into two parts: a finite polymer over a time horizon $T$, which we will call the anticipation time, and the remainder of the cost is integrated by averaging over fluctuations, which in an isolated scenario with no interactions results in a constant force acting on the end of the polymer driving it toward the target. Leads to the question of how do we characterize where can we split it into two parts, how do we link the fluctuations to the parameters and what does "too strong" mean.

%\subsection{Characterization of the polymers in a simple case}
%To better characterize this limit and have a better grasp of the analogy, we present here some analytical results regarding the statistical properties of these polymers in a simple case where the velocity cost is $c(v) = v^2$. This will lead to the introduction of a kind-of Peclet number describing the fluctuations of the polymers, but also characteristic times helping adjusting the numerical parameters, which answers the precedent questions.

\subsection{Discrete-time chains in the very short run}

By contrast, before the anticipation horizon $\TantBare$, correlations are likely to be important. To solve the Langevin-like equation (Eq.~\ref{eq:Langevin1}), we discretize time using a small time step $\mathrm{d}t$ (which will coincide with the decisional update time $\delta t$ in practice). The trajectories $\trajA(t')$ are thus atomized into a chain of successive positions $\boldsymbol{r}_n \hat{=} \trajA(t'=t+n\dd t)$, for $n=0,\dots,N$ ($N \hat{=} \left\lfloor \frac{\TantBare}{dt} \right\rfloor$), so that Eq.~\ref{eq:Langevin1} becomes

\begin{align}
    \frac{\partial \boldsymbol{r}_n}{\partial \tau} = -\frac{\partial \mathcal{C}[\boldsymbol{r}]}{\partial \boldsymbol{r}_n} + \eta(\tau, t+n\dd t).
    \label{eq:Langevin_atomized}
\end{align}
For an isolated agent, remembering that the target cost $c_g$ in Eq.~\ref{eq:integral_cost} has been integrated out into a final cost $\mathcal{C}_f$ acting on atom $N$  and using $v(t+n\dd t)=\frac{\boldsymbol{r}_{n+1}-\boldsymbol{r}_n}{dt}$, the gradient term reads
\begin{align}
    \forall n \in & \left\llbracket 1,N-1\right\rrbracket, \nonumber   \\
   \frac{\partial \mathcal{C}[\boldsymbol{r}]}{\partial \boldsymbol{r}_n}  &=\frac{\alpha \beta}{\dd t^{\beta - 1}}\Big[(\boldsymbol{r}_{n}-\boldsymbol{r}_{n-1})^{\beta-1} - (\boldsymbol{r}_{n+1}-\boldsymbol{r}_{n})^{\beta-1}\Big]
\end{align}
These terms may be interpreted as cohesive forces that hold successive positions close to each other; in the specific case $\beta = 2$, they correspond to linear springs.
The artificial Langevin equation (Eq.~\ref{eq:Langevin_atomized}) therefore describes the dynamics of a thermalized chain of atoms fixed at one end (namely, $\boldsymbol{r}_{0}=\boldsymbol{0}$, in an appropriate reference frame) 
and pulled by the gradient of $\mathcal{C}_f(\boldsymbol{r}_N)$ at the other end; successive atoms are bound by spring forces (because of the velocity cost): this is an active polymer grafted at one end and immersed in a thermal gradient \cite{nguyen2021emergent}. This polymer analogy will be extensively exploited in the next section. Beyond $\TantBare$, our continuum mean-field approach bears some similarity with the mean-field approximation in polymer field theory.

\begin{figure}
    \centering
    \includegraphics[width=1\linewidth]{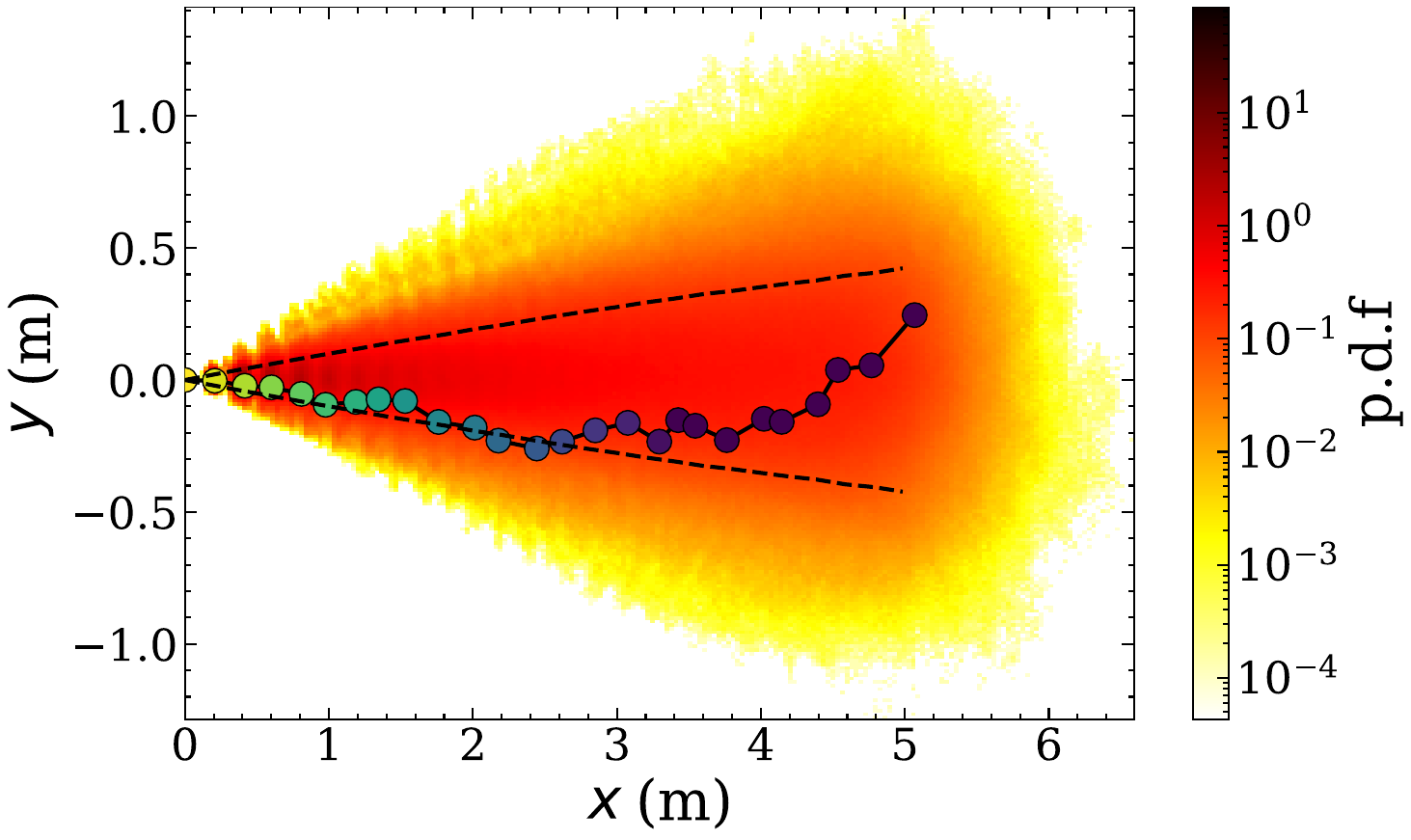}
    \caption{Probability density function of the (thermalized) atoms forming the chain, i.e., the \emph{planned} trajectory, of an isolated purposeful agent starting at $$(x,y)=(0,0)$$ and intent on moving to the right. A snapshot of the chain configuration is depicted with circles, with colors representing the anticipated time from $t'=t$ to $t'=\TantBare$. The black dashed lines are the RMSD along $t'$, as predicted by Eq.~\ref{eq:RMSD}.}
    \label{fig:snapshot-polymere}
\end{figure}

\section{\label{sec:Insights_CondMat}Insights from condensed matter and numerical implementation}

In this mostly technical section, we explain how insights from condensed matter can be leveraged to derive quantities of interest for the anticipatory trajectories and we detail the implementation of the model.

\subsection{Relaxation times of the isolated chain}

\label{sec:condensed-matter}
To get insight into the dynamics of the polymer-like chains, we build on the well-known  Rouse model, taking due account of the specifics of our setting, with polymers fixed at one end and pulled towards the target at the other end. From now on, the exponent $\beta$ in the bio-mechanical cost is set to 2, $c_v(v)=\alpha v^2$, which will be the relevant value for our application to pedestrian crowds. The dynamics in artificial time $\tau$ thus obey:

\begin{subequations}
    \begin{align}
           % \dfrac{\partial \boldsymbol{r}_0}{\partial \tau} &= -\frac{2\alpha}{\dd t}(2\boldsymbol{r}_{0}-\boldsymbol{r}_{1}) + \eta(\tau, t)\\
             \forall n \in & \left\llbracket 1,N-1\right\rrbracket, \nonumber \\
             \ \dfrac{\partial \boldsymbol{r}_{n}}{\partial \tau} &= -\frac{2\alpha}{\dd t}(2\boldsymbol{r}_{n}-\boldsymbol{r}_{n-1} - \boldsymbol{r}_{n+1}) + \eta(\tau, t+n\dd t)\\
            \dfrac{\partial \boldsymbol{r}_{N}}{\partial \tau} &= -\frac{2\alpha}{\dd t}(\boldsymbol{r}_{N}-\boldsymbol{r}_{N-1}) + \eta(\tau, t+\TantBare) + \boldsymbol{F}, \label{eqrN}
    \end{align}
    \label{overdamped_langevin}
\end{subequations}
where we arbitrarily set $\boldsymbol{r}_{0}=\boldsymbol{0}$, out of convenience, and $\boldsymbol{F} \hat{=} \frac{\partial \mathcal{C}_f}{\partial \boldsymbol{r}_n} = \frac{2\kappa}{v_0}\frac{\partial D_g}{\partial \boldsymbol{r}_n}$ is assumed to be constant for the time being, corresponding to a fixed heading direction. In schematic form, using Einstein's summation convention, Eq.~\ref{overdamped_langevin} reads 
$\partial \boldsymbol{r}_{n} / \partial \tau = M_{nm}\boldsymbol{r}_{m}+\dots$. Diagonalizing the matrix $M_{nm}$ gives eigenvalues 
\begin{equation}
\lambda_p = \frac{8\alpha}{\dd t}\sin^2\Big(\frac{2p+1}{2N+1}\frac{\pi}{2}\Big),\ p\in \left\llbracket 0,\,N\right\rrbracket
\end{equation}
% CORRECTION DU 27/05 8\alpha/dt, et non 2\alpha/dt
and (oscillatory) eigenmodes $\boldsymbol{u}_p(\tau)$ such that 
\begin{equation}
    \frac{\partial \boldsymbol{u}_p}{\partial \tau} = -\lambda_p\boldsymbol{u}_p + \tilde{\eta}_p + \alpha_p^{(N)}\boldsymbol{F} ,
    \label{eq:isolated_polymer_transformed}
\end{equation}
where $\tilde{\eta}_p$ is the transform of $\eta_n$ in the $p$-basis. The transformation from the $n$-basis to the $p$-basis is not a simple Fourier transform, because of our particular boundary conditions; it involves coefficients $\alpha_p^{(n)}$ that satisfy 
\begin{equation}
    \boldsymbol{r}_n(\tau)=\sum_{p=0}^{N} \alpha_p^{(n)} \boldsymbol{u}_p(\tau)\textrm{ and }\boldsymbol{u}_p(\tau)=\sum_{n=0}^{N} \alpha_p^{(n)} \boldsymbol{r}_n(\tau).
\end{equation}
Analytical details can be found in Appendix~\ref{app:calc_RMSD}.
An Ornstein-Uhlenbeck process is then obtained by changing variables, $\boldsymbol{u}_p \xrightarrow{}\tilde{\boldsymbol{u}}_p = \boldsymbol{u}_p +  \frac{\alpha_p^{(N)}}{\kappa_p}\boldsymbol{F}$.

\begin{figure}[h!]
  \centering
  \includegraphics[width=0.8\linewidth]{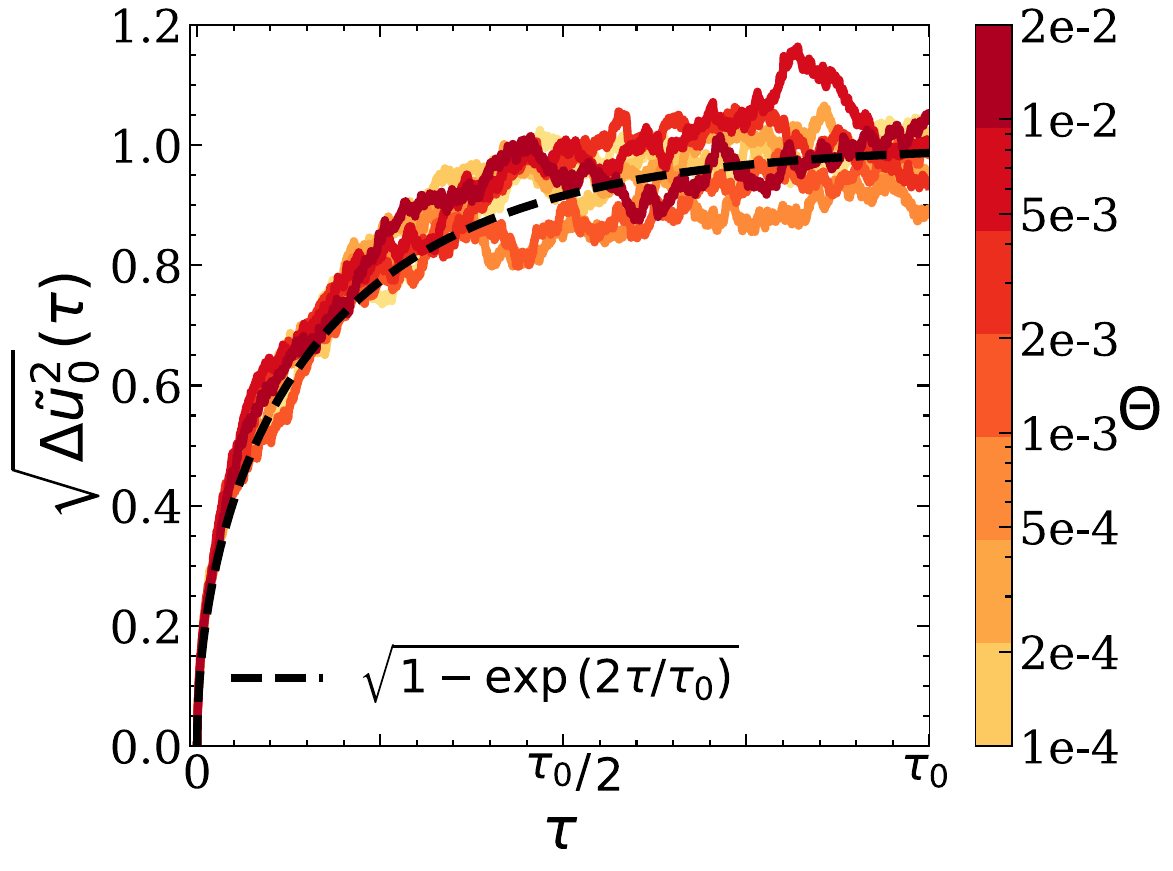}
    \caption{Evolution in fictitious time $\tau$ of the RMSD of the first oscillatory mode of the polymer, $\boldsymbol{u}_0$, for different temperature gradients $\Theta$. The RMSD have been normalized by their predicted values at $\tau=\infty$. The black dashed line is our theoretical prediction, with the characteristic time $\tau_0$ given by Eq.~\ref{eq:tau_0_tau_N}.}
    \label{fig:modes}
\end{figure}

The foregoing equation gives direct insight into the (oscillatory) relaxation modes of the chains, in particular, 
the shortest and longest relaxation times
\begin{equation}
    \tau_0 \approx \frac{2\dd tN^2}{\alpha\pi^2},~  \tau_N \approx \frac{\dd t}{8\alpha}.
    \label{eq:tau_0_tau_N}
\end{equation}

These times govern the choice of a suitable numerical time step, here $\dd \tau = \tau_N/10$, and the duration of the initial `equilibration'. 
Concretely, the system will initially thermalize for $\tau = 5\tau_0$, at $t=0$. After this period, the position $\boldsymbol{r}_1$ of the first atom will be frozen; atom 0 will be destroyed and a new atom will be appended at the chain's end, at $\boldsymbol{r}_N$; and real-time $t$ (and the temperature gradient) will be shifted by $\dd t$, 
following the iterative scheme described in Sec.~\ref{sec:artificial_dynamics} and Fig.~\ref{fig:tubes}; the thermalization will then start over again, adding one atom to the end of the last known polymer configuration. These subsequent thermalizations will be shorter:  $\tau = 5\tau_0\dd t/t_\mathrm{cor}$, drawing on the fact that it takes some (real) time $t_\mathrm{cor}$ for a trajectory plan to drastically change. In terms of algorithmic complexity,  it follows that the number of basic operations to evolve the chain (made of $N$ atoms) by one real time unit  scales with $\mathcal{O}\left( \dd t\,N^3\right)$, that is, $\mathcal{O}\left(\TantBare^3 / \dd t^2\right)$.  

The polymer analogy also points to a caveat in the implementation, which will prove practically relevant and prickly: polymers may get entangled. In other words, in a complex environment, the chains may get trapped in a sub-optimal local trajectory basin during the equilibration, for instance, a trajectory that unrealistically crosses a physical obstacle that cannot be perturbatively circumvented. Empirically, this issue was observed in the crossing of a cluttered environment (Sec.~\ref{seq:cluttered}) when the density was high. When such crossings occur, the polymer is truncated at the $n^\star$-th atom after which the unphysical segment is observed. The atoms $n\geqslant n^\star$ are then repositioned at  $\boldsymbol{r}_{n^\star}$ and thermalization restarts.

\subsection{Fluctuation amplitudes}

\begin{figure}
    \centering
    \includegraphics[width=0.8\linewidth]{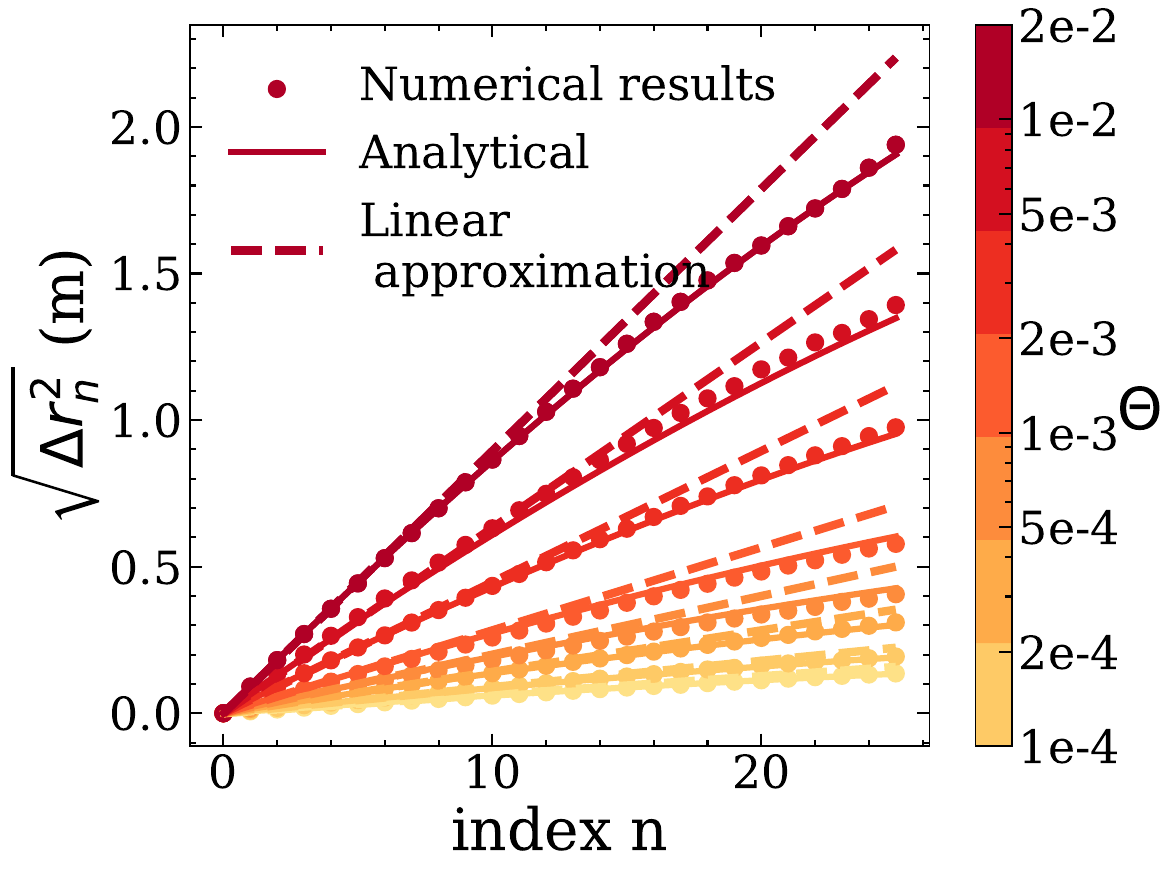}
    \caption{RMSD of the successive atoms $n$ of the chain of an isolated agent in the long-time limit $\tau > \tau_0$, for different temperature gradiens $\Theta$. The linear approximation in dashed line corresponds to Eq.\ref{eq:RMSD}, while the full analytical expression is shown in solid lines. Numerical results have been averaged over 100 realizations.}
    \label{fig:RMSD}
\end{figure}

Following the qualitative arguments of Sec.~\ref{sec:IIB}, the amplitude of the white noise $\eta$ is chosen to increase quadratically with future time $t'$, starting from zero at $t$, viz., 

\begin{equation}
    \langle \eta(\tau,t+n\dd t) \eta(\tau',t+m\dd t)\rangle= 4\Theta (n \dd t)^2\delta_{nm} \delta(\tau-\tau')  .
\end{equation}
    Solving Eq.~\ref{eq:isolated_polymer_transformed} is somewhat more intricate than in the \emph{bona fide} Rouse model, due to our particular boundary conditions, but the mean squared displacement (MSD) of the $n$-th atom ($n\ll N$), derived in Appendix~\ref{app:calc_RMSD}, reduces to a simple approximate expression,
\begin{align}
    \label{eq:RMSD}
    \Delta \boldsymbol{r}_n^2
    &\approx \left(\frac{4}{\pi^2} + \frac{2}{3} \right) \frac{\Theta}{\alpha} N\dd t(n\dd t)^2 \nonumber \\
    & \approx \frac{\Theta \ \TantBare}{\alpha} (t'-t)^2    
\end{align}
The dependence on $\TantBare$ notably stems from the higher thermal agitation at the chain's end segment.
The numerical root mean squared displacement (RMSD) results  shown in Fig.~\ref{fig:RMSD} support the validity of the analytical derivation and (of course to a lesser extent) of its approximation.

The scaling law $\Delta r_n\sim n \dd t$ comforts our choice of a noise that grows at least quadratically in (anticipated) time; otherwise, the \emph{marginal} increase of the uncertainty on future positions would have been decreased with time. %This scaling seems reasonable as it reflects typical uncertainty propagation over time. 

In light of these considerations, we define a dimensionless number, $\PeBare^{-1}$, to gauge how quickly the prediction precision decays with time, by comparing the RMSD $\Delta r_n$ to the characteristic distance  $v_0 n \dd t$ traveled by the agent during  $n\dd t$ :%{\color{red} * Vérifier le 2 et l'absence de  $\alpha$ au dénominateur *}:
\begin{align}
\label{eq:peclet}
    \textit{Pe}^{-1} \hat{=}\frac{\Delta r(t')}{v_0 (t'-t)} =  \frac{\sqrt{\Theta \TantBare/\alpha}}{v_0}.
\end{align}
We call this quantity a bare inverse Péclet number
because it  measures the relative strength of
diffusion \emph{vs.} elasticity in a single polymer
(hence the `bare' qualification, as opposed to the \emph{dressed} number defined subsequently in the presence of interactions between agents). Note that, had we chosen a faster growth of noise with time, the diffusion-over-elasticity ratio would not have been constant along the chain, but would have increased along it.
From the perspective of the agent, $\PeBare^{-1}$ corresponds to the intrinsic positional uncertainty at a 1-meter horizon: the higher $\PeBare^{-1}$, the more uncertain this future position. The parameter $\PeBare$ is thus a cursor that controls the agents' anticipation capabilities, from completely myopic agents ($\PeBare=0$) to omniscient rational agents ($\PeBare\to \infty$).

Besides $\PeBare^{-1}$, two (dimensional) parameters suffice to fully determine the problem, e.g., $\TantBare$ and $v_0$; the above analysis enables us to deduce the other parameters (e.g., $\Theta$, $\alpha$ and $\kappa$) from them.

\subsection{Effect of interactions between chains}
The foregoing reasoning only holds for an isolated chain. For a chain that evolves in a heterogeneous environment or interacts with other chains (i.e., other agents), the cost of Eq.~\ref{eq:integral_cost} is complemented with an interaction cost $c_{\mathrm{int}}$, viz. 
\begin{equation}
 \mathcal{C} = \int_t^{T}\Big(c_v + c_g + c_{\mathrm{int}}\Big)\dd t',
 \label{eq:def_c_int}
\end{equation}
where we have omitted functional dependencies. The RMSD then deviates from Eq.~\ref{eq:RMSD}. For example, we expect smaller fluctuations (hence, a smaller \emph{dressed} inverse Péclet number $\PeDressed^{-1}$) under strong spatial confinement, but larger ones (hence, a larger $\PeDressed^{-1}$) when the chain is subject to multiple interactions with its neighbors. This suggests a reduced anticipation horizon at higher densities, which we shall confirm numerically in Sec.~\ref{sub:fund_diagram} when the model is applied to pedestrians.

\begin{figure}
    \centering
    \includegraphics[width=1\linewidth]{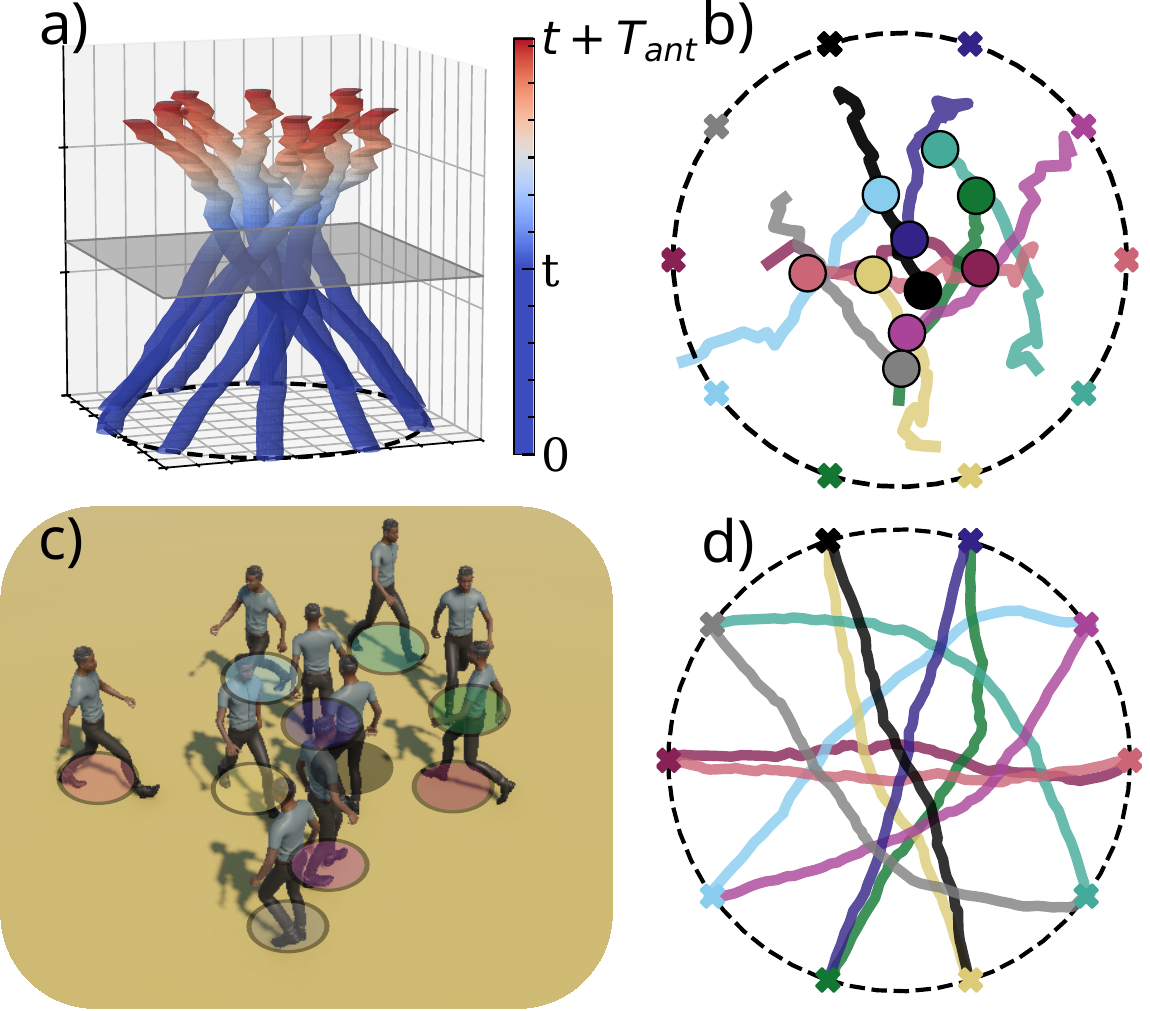}
    \caption{Simulated pedestrian trajectories in an `antipodal' scenario, in which agents are initially on a circle of diameter $\SI{5}{\meter}$ and each one aims for the antipodal position. The pedestrian model is described in Sec.~\ref{sec:level4}. {\bf(a)} Trajectories in space–time. Colors indicate the  random fluctuation intensity; the plane in gray, at present time $t' = t$, separates the future from the past frozen trajectories. {\bf (b)} Corresponding positions (circles) at $t'=t$ and planned trajectories for $t'\geq t$, in space. Destinations are shown as crosses. {\bf(c)} Snapshot of a 3D animation of the dynamics, created in \emph{Unity}; the colors of the shaded circles coincide with those of the 2D trajectories. {\bf(d)} Full trajectories, obtained by simulating the dynamics for $t = \SI{8}{\second}$.}
    \label{fig:antipodal}
\end{figure}
That being said, owing to the analogy with dilute polymer solutions,  the Rouse characteristic times $\tau_0$ and $\tau_N$ introduced above  are expected to still govern the relaxation of individual chains, as long as entanglements remain scarce and interactions are not too nonlinear. Accordingly, we will keep the equilibration times defined above as a function of $\tau_0$ when we simulate interacting chains. 

Additional numerical concerns emerge in this interacting context. Overlaps between chains (i.e., prospective "collisions" between agents) should generally be  prohibited, but may go numerically unnoticed when chains are discretized and leave some space between successive atoms. To avoid these spurious tunneling effects, interactions between chains will be evaluated as a function not of the distance between their respective atoms  $\boldsymbol{r}_n$, but of the minimal distance between corresponding segments $[\boldsymbol{r}_n,\boldsymbol{r}_{n+1}]$, which can readily be obtained numerically. The same method is applied to interactions with the environment (walls in the pedestrian case).

%Déja expliqué dans la fin du IVA sur l'analogie avec les polymeres Secondly, similarly to entangled polymers, the initial configurations of the chains, coarsely slung between the origin and the destination, may be improbable (e.g., going through a wall) and chains may fail to spontaneously escape from these quagmires via the artificial dynamics, despite their high cost. We remedy this by cutting and trimming loose ends: whenever a high-cost segment {\color{red} * préciser * } is detected for longer than $\tau_N$ along a chain, this segment is artificially cut and all atoms on the \emph{future} side are relocated to the position of the last atom before the high-cost segment.

\subsection{Mean-field approach beyond $\TantBare$}
\label{sec:meanfield}
The reasoning developed in Sec.~\ref{sec:beyond-anticipation} to approximate the cost beyond $\TantBare$ can be generalized in the presence of interactions between chains as well as other forms of spatial dependence.

Beyond $\TantBare$, in a mean-field spirit, we want to average interactions along the chains over realizations, and to coarse-grain individual trajectories $\trajA$ into a continuous a density field $\rho(\boldsymbol{r})$. This could be achieved using the framework of mean-field game theory \cite{bonnemain2023pedestrians,butano2024discounted}, but here we opt for a much coarser, but simpler approach:  we consider the fictitious-time-averaged positions of the ends of the other agents' chains and insert a Gaussian halo at this position into the density field $\rho(\boldsymbol{r})$, blurring the halo all the more as the agent is far. More precisely, the standard deviation of the Gaussian with the expected time of interaction (as inferred from the effective distance) follows the RMSD law of Eq.~\ref{eq:RMSD}. Interaction costs at a given prospective position $\boldsymbol{r}$ are then weighted by $\rho(\boldsymbol{r})$, defining an effective spatial potential $\Phi_j(\boldsymbol{r})$, which represents agent $j$’s estimated cost for being at $\boldsymbol{r}$ in the future. A detailed derivation of this construction will be presented elsewhere.

Given a spatial cost field $\Phi(\boldsymbol{r})$ — which can include target attraction, interactions, local discomfort, and environmental heterogeneities — the final cost functional $\mathcal{C}_f$ can be expressed as
\begin{align}
\mathcal{C}_f = \int_{\TantBare}^{T} \left( \alpha v(t')^2 + \Phi(\boldsymbol{r}(t')) \right)\mathrm{d}t'.
\label{eq:effective_MF}
\end{align}
Provided that the positions at the boundaries are fixed, namely, 
$\trajA(\TantBare)=\boldsymbol{r}_N$ and $\trajA(T)=\state_g$ (where $\state_g$ is the location of the goal), optimizing the cost $\mathcal{C}_f$ in Eq.~\ref{eq:effective_MF} comes down to extremizing the action of an inertial particle of mass $2\alpha$ in a potential field $-\Phi$. Thus, the Hamiltonian is constant along the optimal trajectory,
\begin{align}
H(t') = \alpha v(t')^2 - \Phi(\boldsymbol{r}(t')) = \Phi(\state_g) \hat{=}0,
\end{align}
and the optimal speed reads
\begin{align}
v_\mathrm{opt}(\boldsymbol{r}) = \sqrt{\frac{\Phi(\boldsymbol{r})}{\alpha}}.
\end{align}
This result generalizes the expression used in Sec.~\ref{sec:beyond-anticipation}, which corresponds to the specific choice $\Phi(\boldsymbol{r}) = \kappa\big(1 - \chi_{\state_g}(\boldsymbol{r}\big)\big)$.

Changing integration variables $\mathrm{d}t = \mathrm{d}r / v_\mathrm{opt}(\boldsymbol{r})$ in Eq.~\ref{eq:effective_MF} yields a final cost
\begin{align}
\mathcal{C}_f =  \int_{\boldsymbol{r}_N}^{\state_g} 2\sqrt{\Phi(\boldsymbol{r})}\,\mathrm{d}r.
\end{align}
Numerically, $\mathcal{C}_f$ is computed for any initial position $\boldsymbol{r}_N$, using the Fast Marching Method, which propagates a wavefront from $\state_g$  with an effective refractive index $2\sqrt{\Phi(\boldsymbol{r})}$ following the Eikonal equation; then, its gradient is readily obtained.

In passing, a remark about the negative sign in the potential $-\Phi$ is worthy of interest: While typical models for collision avoidance between pedestrians or robots \cite{helbing1995social,van2008reciprocal} posit repulsive interactions, the optimal trajectory here is that of a Newtonian particle interacting \emph{attractively} with its neighbors. However, direct resolution of Newton's equations will obviously go astray, if one is not able to prescribe the initial position and velocity of the particle with exquisite precision.

% A.N.: J'ai fait, ci-dessus, une proposition qui court-circuite l'aspect contrôle optimal et de résoudre le problème en voyant $C_f$ comme l'action d'une part newtonienne dans un potentiel -$\Phi$ (cf dernier paragraphe)-- je ne vois pas vraiment ce qu'apporte la théorie du contrôle ici.

\section{\label{sec:level4}Application to Pedestrian crowds}

\subsection{Context of pedestrian dynamics}

This section is concerned with the application of our anticipatory framework to the modeling of crowd dynamics.
Over the past years, evidence has been amassed that pedestrians exhibit distinctive anticipatory skills, not only at the higher level of route choice, but also in their local navigation in various scenarios, from head-on encounters \cite{murakami2021mutual} to passing through a static crowd \cite{nicolas2019mechanical,bonnemain2023pedestrians} or circumnavigating small groups \cite{bruneau2015going}.
Agent-based models for pedestrian dynamics typically account for such anticipatory capabilities (perceived through the lens of collision avoidance) by supplementing mostly \emph{reactive} equations of motion with specific forces to anticipate collisions \cite{helbing1995social,Karamouzas2014universal,lu2020pedestrian,xu2021anticipation,hu2023anticipation} or constraints aimed at guaranteeing the absence of collisions \cite{van2008reciprocal,van2011reciprocal}. Most of the time, these corrections rely on 
a linear extrapolation of trajectories, i.e., on the assumption that current velocities are conserved \cite{xu2021anticipation, hu2023anticipation, echeverria2025near}. Thus, they are bound to fail in situations requiring more complex anticipatory behaviors, such as pedestrians crossing cluttered environments \cite{raulin2025highs}, or in the scenario sketched in Fig.~\ref{fig:narrow_corridor}, where one agent needs to stop in a side niche and let another one pass. 

Applying our broader anticipatory framework to pedestrians will both clarify some relatively fuzzy notions in the field, such as the decomposition of the dynamics into strategic, tactical and operational \cite{hoogendoorn2005pedestrian}, and remedy many of the aforementioned deficiencies. Remarkably, we will see that even the fairly straightforward implementation of the framework that we put forward outperforms established models notably in a scenario involving crossing static crowds. The proposed model bears some resemblance with the (already discussed, see Sec.~\ref{sub:connections}) adaptation of the CHOMP motion planning algorithm to pedestrian dynamics \cite{ziebart2009planning} (but here pedestrians are duly considered as autonomous agents) and, to a greater extent, to Modi et al.'s pioneering work on handling pedestrian trajectories as spacetime tubes \cite{modi2023mutiagent} (but these researchers viewed the problem as an optimization problem under constraints, rather than delving into anticipation and the uncertainty that comes along with it). There is also some connection with machine-learning based models, in which future pedestrian moves are probabilistically predicted with a neural network, to guide decisions of motion \cite{zhi2021anticipatory}.

\subsection{Cost and Specific Features}

Our starting point is that pedestrians strive to reach their goal (`target') as soon as possible, taking into account penalties incurred along their paths. Put in mathematical terms, each agent $j$ strives to minimize a generalized travel time
\begin{equation}
    TT_j:=\int_t^T c_j(t')dt', 
\end{equation}
where the running cost $c_j(t')$ should vanish when the agent has reached the target area and $T$ is a large upper time bound. Introducing the  complementary indicator function $\bar\chi_j$ (equal to 0 if the agent is in the target zone, 1 otherwise), the local running cost will thus comprise a penalty due to the built environment ($\kappa\,\bar\chi_j (\boldsymbol{r}_j(t'))$), the biomechanical expenditure due to the instantaneous speed ($c_{bm}(||\boldsymbol{\dot{r}}_j(t')||)$ and  penalties $c_\mathrm{int}$ due to contact and to proxemic interactions with the neighbors, as follows
\begin{equation}
    c_j(t')= \kappa\,\bar\chi_j(\boldsymbol{r}_j(t'))  + c_{bm}(||\boldsymbol{\dot{r}}_j(t')||)  + c_\mathrm{int}(\boldsymbol{r}_{1}(t'),\boldsymbol{r}_2(t'),\dots).
    \label{eq:cost_for_peds}
\end{equation}
 For an isolated agent walking at constant speed in uniform space, we notice that $TT_j$ is simply proportional to the time to reach the target area, as expected. In the following, we detail and discretize in time the different contributions.

\subsubsection{Target cost}
For numerical purposes, it is convenient to smooth the (discontinuous) $\bar\chi_j$ in space, viz.
\begin{equation}
    \bar\chi_j (\boldsymbol{r})\simeq 1 - \exp{\Big(-\frac{|\boldsymbol{r}-\state_g|^2}{2\lambda^2}\Big)},
\end{equation}
where $\lambda$ should be of the order of magnitude of the smallest typical pedestrian length scale set to $\SI{0.1}{\meter}$. The target cost enters the dynamics both through the final cost, which drives the chains toward the target, and through the gradient of the Gaussian well, which stabilizes the atoms in the vicinity of the target once it has been reached

\subsubsection{Bio-mechanical cost}
This cost represents the energetic expenditure associated with maintaining a walking speed $v$. Experimentally, by measuring oxygen consumption, it was found to increase quadratically with speed \cite{ludlow2016energy}, 
\begin{equation}
    c_{bm}(v) =cst+ \alpha v^2.
    \label{eq:biomech_cost}
\end{equation}
The chosen parameters $\alpha$ and $\kappa$ control the preferred speed of the pedestrian. As described in Eq.~\ref{eq:vitesse} for $\beta = 2$, $v_\mathrm{opt} = \sqrt{\kappa/\alpha}$ where $v_\mathrm{opt}$ is a given parameter that depends on the pedestrian, usually ranging from \SI{1}{\meter\per\second} to $\SI{1.5}{\meter\per\second}$. Since the total cost is defined up to a multiplicative constant, $\alpha$ can be set to $\frac{1}{2}$ and $\kappa$ is deduced from $v_\mathrm{opt}$.

%the form $A+Bv^2$, with a constant term penalizing the initiation of motion and a quadratic dependence on velocity. The constant term introduces a discontinuity between $v=0$ and $v>0$, which is inconvenient for implementation in a gradient descent framework. Therefore, the expression of the energetic cost is modified for %$v<v^*=\SI{0.5}{\meter\per\second}$ to smooth this discontinuity:

%The polynomial expression for $v<v^*$ has been chosen to ensure the continuity of $c_v$ and its derivative at $v=v^*$. The choice of the value $v^* = \SI{0.5}{\meter\per\second}$ comes from the analysis of several speed histograms, revealing that very few pedestrians move comfortably below this threshold. In other words, a pedestrian generally prefers either to remain stationary or to move at a speed at least equal to $v^*$, a behavior likely related to the discomfort associated with steps that are too short or too slow.

\subsubsection{Perception and interactions between agents }

 Following common practice in pedestrian dynamics, the interactions between agents are divided into two contributions, contact interactions and social ones:
\begin{equation}
    c_\mathrm{int}^{j\leftarrow i}(s_{ij})= c_{\mathrm{contact}}^{j\leftarrow i}(s_{ij})
    +  c_{\mathrm{social}}^{j\leftarrow i}(s_{ij}). 
\end{equation}
where the interaction depends on the anticipated spacing $s_{ij}=||\boldsymbol{r}_i(t')-\boldsymbol{r}_j(t')||-R_i-R_j$ between the agents, of radii $R_i$ and $R_j$, at time $t'$. Here, additive pair-wise interactions have been assumed, as the total interactions cost is summed over $j$. In practice the quick spatial decay of $c_\mathrm{int}^{j\leftarrow i}$ will ensure that the closest agents dominate the interaction.

The contact contribution, the first term, is activated only when the spacing $s_{ij}$ is negative, hence the use of a Heaviside function $\Theta$, and it is modeled as a soft-sphere contact between supposedly deformable human bodies:
\begin{equation}
    c_{\mathrm{contact}}^{j\leftarrow i}(s_{ij}) = b_{\mathrm{contact}}\Theta(-s_{ij})\,s_{ij}^2.
\end{equation}

For the second term, the (anticipated) social interactions between an agent $j$ and other agents $i$ are affected by how $j$ perceives them and anticipates their evolution through space. This was integrated into a perception \emph{warping} function $\pi^j_{t'}$ in the general context of Sec.~\ref{sec:shared-space}. Here, to keep the model as simple as possible, we reduce these considerations to whether other agents $i$ are, or not, in the visual cone of $j$  at present time $t$ and accordingly we modulate the interaction strengths with  a smooth function $\cos^{+}(\theta_{ij}) = \max{\left(0,\cos\,\theta_{ij}(t)\right)}$ of the angle between agent $j$'s current direction and agent $i$ current position relative to $j$:

\begin{align}
c_{\mathrm{social}}^{j\leftarrow i}(s_{ij})=
b_{\mathrm{social}}\,\cos^{+}\left(\theta_{ij}\right)  \,\exp\left(\frac{-s_{ij}}{\sigma_{ij}}\right).
\end{align}
This term represents the longer-range proxemic  interaction
\cite{hall1969hidden,helbing1995social}, whereby an agent is intent on preserving their `social bubble' of characteristic size $\sigma_{ij}$.

The coefficients $b_{\mathrm{social}}$ and
$b_{\mathrm{contact}}$ should satisfy: $0<b_{\mathrm{social}} \ll b_{\mathrm{contact}}\delta_r^2$ (with $\delta_r \ll R$ ); they
have been adjusted by considering a schematic scenario of face-to-face avoidance.

\subsubsection{Obstacle avoidance cost}

Prohibited regions in the environment (walls, obstacles, ...) impact the dynamics of chains in the same way as contacts with other pedestrians, i.e., via soft-sphere interactions, but, for computational efficiency, these costs and their gradients are pre-computed on a fine-mesh lattice: obstacles are defined on a lattice and at every lattice node the distance to the nearest obstacle is (pre)-computed using the Fast Marching algorithm described in Sec.~\ref{sec:meanfield}.
%{\color{red} Vérifier qu'il n'y a pas d'interactions "social" => OK} Ok !

\begin{table}[t]
\centering
\caption{Parameters of the model applied to pedestrian dynamics.}
\begin{tabular}{lcl}
\toprule
Symbol & Definition & Value \\
\midrule

\multicolumn{3}{c}{\textit{Agent parameters}} \\
\midrule
$R$ 
& radius 
& $[0.15, 0.25]~\SI{}{\meter}$ \\

$v_0$ 
& Preferred speed 
& $[1, 1.5]~\SI{}{\meter\per\second}$ \\

$\alpha$ 
& Biomechanical prefactor 
& $1/2$ \\

$\kappa$ 
& Attraction toward the target 
& $\alpha v_0^2$ \\
$\lambda$ 
& Width of target well 
& $\SI{0.1}{\meter}$ \\
\midrule
\multicolumn{3}{c}{\textit{Interaction parameters}} \\
\midrule

$\sigma$ 
& Social interaction range 
& $\SI{0.2}{\meter}$ \\

$b_{\mathrm{social}}$ 
& Social repulsion coefficient 
& $1$ \\

$b_{\mathrm{contact}}$ 
& Contact repulsion coefficient 
& $\SI{300}{\per\meter\squared}$ \\

$b_{\mathrm{obstacles}}$ 
& Obstacle repulsion coefficient 
& $\SI{300}{\per\meter\squared}$ \\

\midrule
\multicolumn{3}{c}{\textit{Simulation parameters}} \\
\midrule

$\dd t$ 
& Time step 
& $[0.1, 0.2]~\SI{}{\second}$ \\

$\textit{Pe}^{-1}$ 
& Bare inverse Péclet number 
& $[0.05,\,0.1]$ \\

 $\TantBare$
 & Chain length (in time) 
 &$[2,\,8]~\SI{}{\second}$ \\
 
$\dd \tau$ 
& Fictitious time step
&  $\dd t / 80 \alpha$ \\

\bottomrule
\end{tabular}
\label{tab:parameters}
\end{table}

\subsection{Fundamental diagrams}
\label{sub:fund_diagram}
A key quantity in applied crowd dynamics \cite{zhang2011transitions, helbing2007dynamics, weidmann1993transporttechnik, Vanumu2017Dec},  the fundamental diagram relates the average speed or flow of a crowd to the pedestrian density in given settings (this is the counterpart of constitutive equations in Mechanics). The speed is approximately constant in the low-density, free-flow regime, and then decreases with density. This leads to an increase of non-monotonic fundamental diagram for the flow, which increases at relatively low densities, before reaching a peak and falling off.

Let us consider the fundamental diagram in unidirectional and bidirectional flows in a corridor of width $\SI{2.5}{\meter}$ (i.e., wide enough to have several pedestrians walk side by side) and (periodic)  length $\SI{10}{\meter}$. These scenarios were simulated with a crowd of agents of non-uniform radii, randomly drawn from a normal distribution with mean $R = \SI{0.22}{\meter}$ and standard deviation $\SI{0.02}{\meter}$. For compatibility with the periodic boundaries along the longitudinal axis, the agents' targets are constantly relocated $\SI{4}{\meter}$ in front of them, like a carrot in front of a donkey.  
\begin{figure}[h!]
    \centering
    \includegraphics[width=0.9\linewidth]{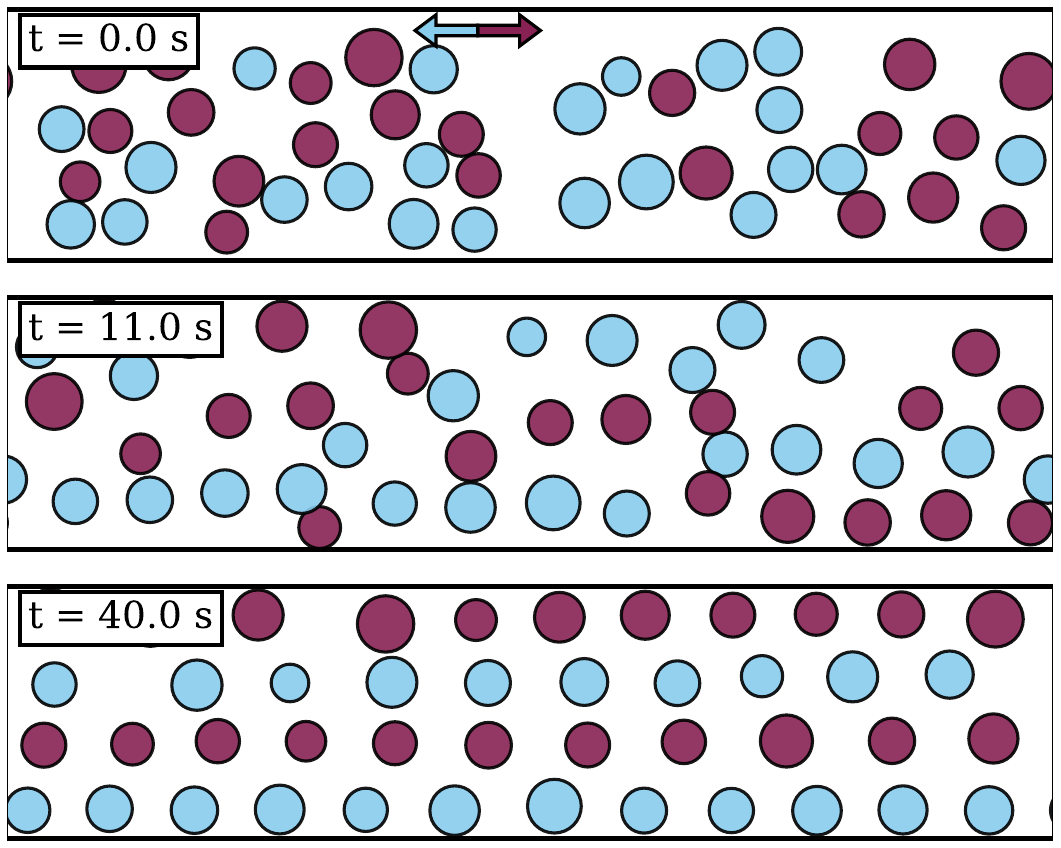}
    \caption{Lane formation in a bidirectional corridor flow ($\PeBare^{-1} = 0.05$): snapshots at $t=0\,\mathrm{s},\ 11\,\mathrm{s},\ 40\,\mathrm{s}$. The corridor is $\SI{10}{\meter}$ long in the horizontal (periodic) direction and $\SI{2.5}{\meter}$ wide; the density is $\rho = \SI{1.8}{\mathrm{Ped\per\meter^2}}$ (44 agents).}
    \label{fig:snapshot_bidir}
\end{figure}

Snapshots of the simulations are shown in Fig.~\ref{fig:snapshot_bidir}, at different times, from the random initial configuration to the steady state. In the bidirectional flow conditions, we observe the formation of lanes; this  well documented phenomenon is present in a variety of related systems, such as driven binary mixtures \cite{Poncet2017Mar, Dzubiella2002Jan}, as well as in the basic social force model \cite{helbing1995social}. However, it is worth noticing that in our simulations lanes are not always stable, in particular for densities close to the critical density (beyond which the crowd does not move). Thus, the \textit{steady state} is not truly stationary.  Besides, the characteristic time required for lane formation increases as the system approaches the critical density, suggesting a slowing-down behavior reminiscent of the divergence of the characteristic times at phase transitions in binary active liquids \cite{Bain2017Jun}.

\begin{figure}[h!]
    \centering
    \includegraphics[width=0.85\linewidth]{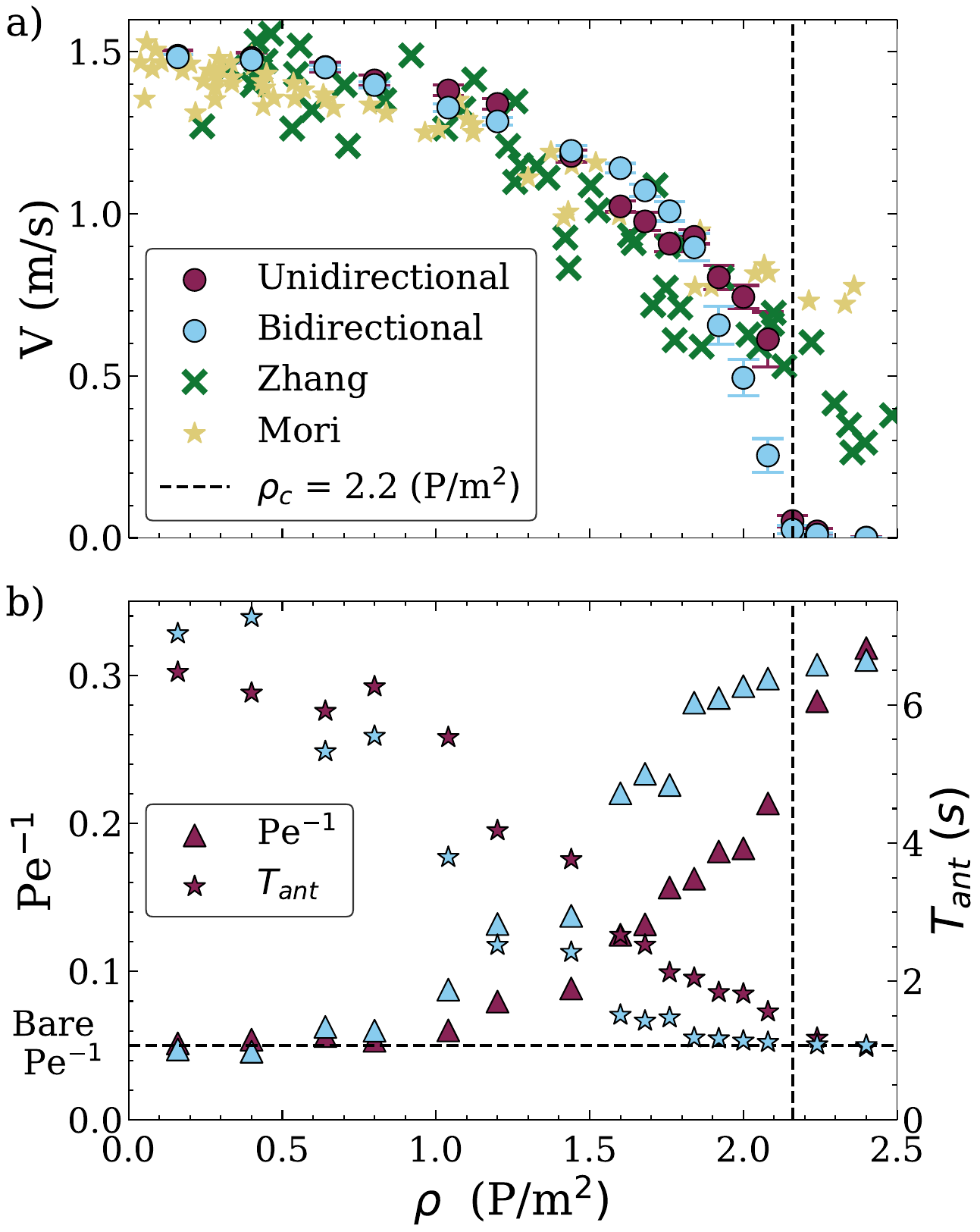}
    \caption{Unidirectional and bidirectional corridor flows. {\bf(a)}  Fundamental diagram, relating the average longitudinal speed to the pedestrian density. Error bars:  standard error, among 10 realizations for the bidirectional case and 2 realizations for the unidirectional case. Experimental measures from \citeauthor{zhang2011transitions} \cite{zhang2011transitions} and \citeauthor{Mori1987May} \cite{Mori1987May} are shown for comparison. {\bf(b)} Dressed $\PeDressed^{-1}$ (computed from the RMSD measured during equilibration) and associated effective anticipation time $\TantEff$ obtained via Eq.~\ref{eq:dressedpeclet}, as a function of density. Refer to Fig.~\ref{fig:snapshot_bidir} for details about the corridor geometry.
    The bare Péclet number was fixed at $\PeBare^{-1} = 0.05$.}
    \label{fig:fund_diagram}
\end{figure}

The fundamental diagrams presented Fig.~\ref{fig:fund_diagram}a are obtained by running such simulations at various densities and measuring the average longitudinal speed after $\SI{50}{\second}$.
In the low-density region, they agree well with experimental measurements \cite{zhang2011transitions, Mori1987May}, without any \emph{ad hoc} adjustment. They reproduce the gradual decay from the free-flowing speed (as do many other pedestrian models). Then the mean speed is found to plummet to zero at a critical density $\rho_c = \SI{2.2}{\mathrm{Ped}\per\meter^2}$, with large error bars that echo the aforementioned absence of a truly stationary steady state in this region. This critical density is much lower than densities at which crowds are still found to move experimentally and empirically (\SIrange{4}{8}{\mathrm{Ped}\per\meter^2}) \cite{Vanumu2017Dec}. As in other models \cite{echeverria2023body}, this is most probably due to the circular shape of the agents, which enhances surface coverage, hence excluded-volume effects, and also bars torso rotations and other anisotropic behaviors whereby real pedestrians can temporarily reduce their effective cross section.

Figure~\ref{fig:fund_diagram}b presents a more original perspective on these corridor-flow experiments, by showing how the \emph{effective} anticipation horizon $\TantEff$, defined by t $\Delta r(t+\TantEff) \simeq \ell$ (where $\ell= \SI{0.5}{\meter}$ is a characteristic length), evolves with density. In practice, we first measure the  polymer fluctuations at $t=0$ during the Langevin dynamics, deduce the dressed $\PeDressed^{-1}$ from Eq.~\ref{eq:peclet}, and then find $\TantEff$ from the relation:
\begin{align}
    \label{eq:dressedpeclet}
   \ell=\Delta r(t+\TantEff) = \PeDressed^{-1}\,v_0\,\TantEff.
\end{align}
In Fig.~\ref{fig:fund_diagram}b, we see that the dressed inverse Péclet number $\PeDressed^{-1}$  is equal to its bare counterpart (Eq.~\ref{eq:peclet}) at vanishing density, but then grows with density, while the anticipation time concomitantly decreases from approximately $\TantEff=\SI{7}{\second}$ to $\TantEff=\SI{1}{\second}$. These evolutions are caused by enhanced fluctuations of the chains, which we explain as follows: As congestion increases, possible paths around neighbors start to proliferate and individual trajectories become less predictable.

\subsection{Constrained navigation and cluttered environments}
\label{seq:cluttered}
Properly modeling anticipatory dynamics plays a more important part in the complex scenarios with multiple interacting agents or cluttered environments that we probe in this section.

\subsubsection{Antipodal configuration}
First consider a scenario \cite{Xiao2019Jun}, in which pedestrians are initially spaced along a circle and are asked to walk to the antipodal position. This antipodal scenario is known to raise intense conflicts between agents at the center, hence issues for models. Even when models avoid deadlocks thanks to the integration of some level of anticipation \cite{van2008reciprocal, van2011reciprocal, Karamouzas2014universal}, they often produce trajectories that deviate from a straight line (along the diameter) only quite late, or that are highly symmetric between agents, as though the collective motion has been harmoniously organized. By contrast, the trajectories generated with our model, displayed in Fig.~\ref{fig:antipodal}d, 
feature detours that are undertaken quite early, as soon as the agents leave their initial positions, in anticipation of the congestion that will arise in the central zone.

To give an estimate of the computational cost, the simulation of the 10-pedestrian scenario takes slightly longer than real time ($\approx\SI{10}{\second}$) on a CPU. The computational cost, however, scales with the square of the number of pedestrians due to the increasing number of pairwise interactions, and increases further when the time step $\dd t$ is reduced to achieve a more accurate description of contact interactions (as requried to describe the metro exit in Sec.~\ref{sec:metroexit}). At this stage, we have not made substantial efforts to optimize or parallelize the code; still, all simulations presented in this article ran in reasonable times, typically within a few hours. 

\begin{figure}[h!]
    \centering
    \includegraphics[width=1\linewidth]{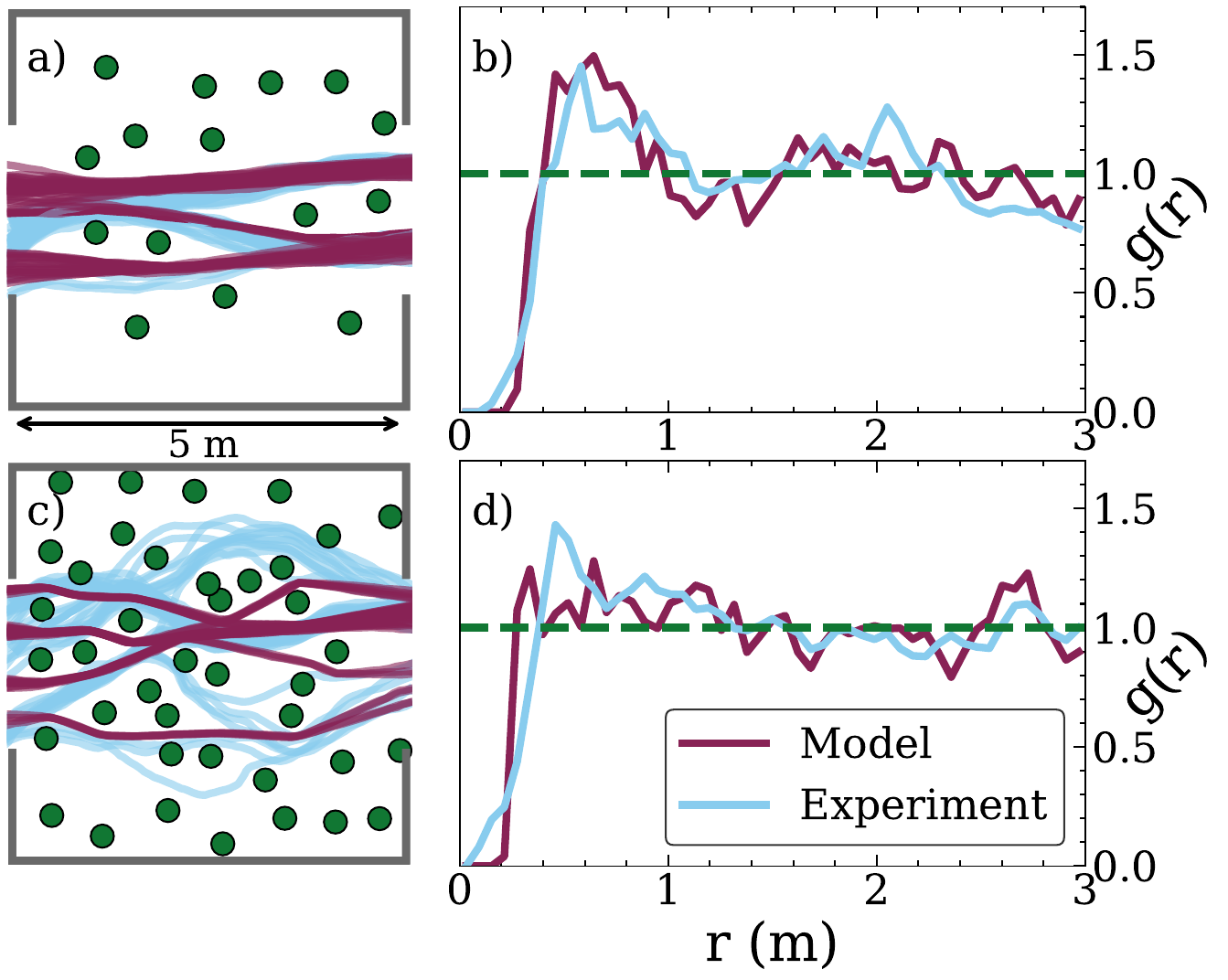}
    \caption{Crossing a static crowd. {\bf(a, c)} Experimental trajectories measured  by Wang et al. \cite{wang2023exploring} (\emph{in blue}) and their numerical counterparts (\emph{in red}) at (a) low density $\rho = \SI{0.6}{\mathrm{Ped\per\meter^2}}$  and (c) high density $\rho = \SI{1.6}{\mathrm{Ped\per\meter^2}}$. {\bf(b, d)} Pair distribution functions (pdf) $g(r)$ between the crossing pedestrian and the static ones, computed from experimental data (\emph{in blue}) and numerical simulations (\emph{in red}). $g(r)=G(r)/G_{NI}(r)$ was normalized by $G_{NI}(r)$, the pdf obtained using the same static configuration, but crossing trajectories from uncorrelated trials. The baseline $g(r)=1$ is shown in green. Simulations were performed at $\PeBare^{-1} = 0.1$, with a pedestrian radius $r = \SI{17}{\centi\meter}$. Results for other densities are displayed in Appendix~\ref{appendix:wang}, Fig.~\ref{fig:wang_full_traj},\ref{fig:wang_full_proba}}
    \label{fig:crossing}
\end{figure}

\subsubsection{Crossing of a static crowd}
A second, more complex scenario consists in crossing a cluttered space.  Wang et al \cite{wang2023exploring} investigated this scenario experimentally by asking participants to cross one-by-one a confined area cluttered with static pedestrians as obstacles.  These obstacles will be modeled as non-moving agents, standing at the same positions as in the experiments. In \cite{raulin2025highs}, we showed that prominent models for pedestrian dynamics utterly fail to capture the crossing motion, above some (modest) density of static agents, even if the agents' radii are set to barely 10~cm; many agents end up in `dead ends' because of too short-sighted anticipation.

To simulate these scenarios, we initialize the crossing agents at the same initial positions as participants in the experiments (on the right-hand side in Fig.~\ref{fig:crossing}a,c), and we set their target as a destination zone located on the opposite (left-hand) side of the confined area. The simulated trajectories for a sparse static crowd and for a denser one are shown Fig.~\ref{fig:crossing}a,c and compared to the experimental ones. Contrary to what we had found with existing models \cite{raulin2025highs}, all simulated agents manage to cross the crowd, even in the denser case. Furthermore, on the whole, the simulated trajectories are similar to the experimental observations. Note, however, that the \emph{experimental} trajectories (but not the simulated ones) tend to bend downwards close to the exit, on the left, probably because the instruction given to participants to resume their starting positions introduces a bias. 

For a finer-scale comparison centered on interpersonal distances, we plot the radial pair distributions $g(r)$ between the crossing pedestrian and the static individuals in Fig.~\ref{fig:crossing}b,d; $g(r) \hat{=} G(r)/G_{NI}(r)$ was normalized by the pair distribution $G_{NI}(r)$ obtained using the same static crowd configuration, but crossing trajectories from other realizations, so that the baseline $g(r)=1$ corresponds to situation without interactions. Once again, model and experiments are found to agree relatively well, with a dip in the vicinity of $r=0$ due to short-range repulsions, followed by a peak. The rise from $r=0$ is smoother in the experiments (where the static pedestrians are of course not circular), whereas  in the model they rise more sharply at $r=\SI{17}{\centi\meter}$ because of the soft-sphere contact interactions which strictly prevent overlaps. Nonetheless, the results show that, even though our focus was on anticipation and we only used very simple interaction terms between circular objects in the cost function, the model accurately captures pedestrian behavior at the individual scale.
Similar results are obtained for the other scenarios tested in \cite{wang2023exploring} with static crowds, as shown in Appendix~\ref{appendix:wang}. We also confirmed that the model gives realistic results in the scenarios in which the crowd is not static, but moving (\emph{not shown}).

\subsubsection{Anticipation in a narrow corridor}
\begin{figure}[h!]
    \centering
    \includegraphics[width=1\linewidth]{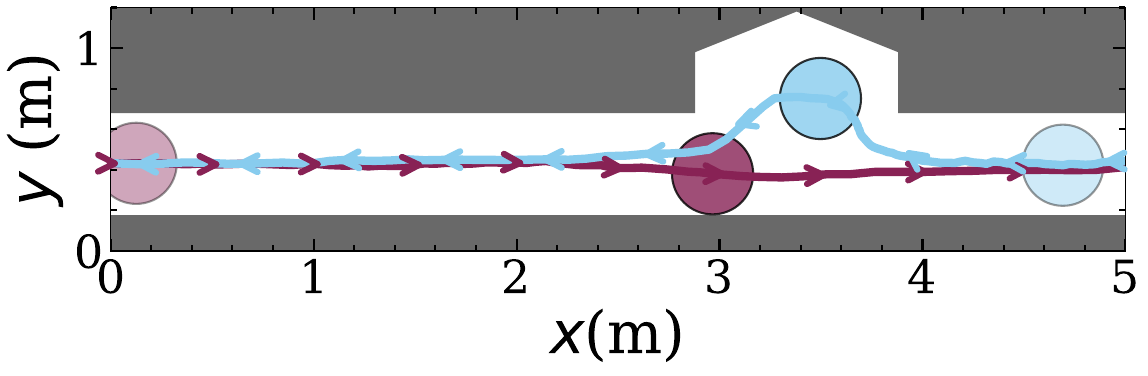}
    \caption{Agents walking in opposite directions in a one-person-wide corridor with a side niche ($\PeDressed^{-1} = 0.1$). The agents' positions at $t= \SI{0}{\second}$ (\emph{transparent disks}) and $t = \SI{5}{\second}$ (\emph{opaque disks}) are displayed; the full trajectories are shown as lines.}
    \label{fig:narrow_corridor}
\end{figure}

Thirdly, we investigate a scenario in which anticipation may be even more critical, because of geometric constraints: two pedestrians have to pass one another in a very narrow corridor with a small niche on one side (see Fig.~\ref{fig:narrow_corridor}). One pedestrian needs to anticipate sufficiently far ahead to understand that the most only viable strategy is to wait in the niche and let their counterpart pass, before moving forward. This nontrivial strategy is successfully adopted by the simulated agent (in blue in Fig.~\ref{fig:narrow_corridor}), without prescribing any additional rule to enforce yielding.

\subsection{Tactical route choice}
\label{sub:tactics}
Pedestrian dynamics usually makes the distinction between three levels of path planning: strategic, tactical, and operational, each of them acting on a different time and length scale. The strategic level defines the macroscopic framework of crowd organization, including large-scale decisions such as scheduling, global route design, and transportation logistics. At the tactical level, the focus shifts to intermediate-scale processes, namely route choice and collective guidance mechanisms. Finally, the operational level describes the microscopic dynamics, addressing individual motion and local interactions within the crowd. 

Our anticipatory framework shines a new light on this decomposition \cite{echeverria2025near}, through the lens of uncertainty. One transits from the operational to the tactical level when agents can no longer anticipate localized trajectories but make decisions based on averaged density fields, i.e., at the effective anticipation horizon $\TantEff$. At even higher levels of uncertainty (not considered in this paper), the very geometry of the environment gets blurred and specific routes can no longer be distinguished; in other words, temperature is so high that the polymer chains may cross walls. This marks the entrance into the strategic level, where choices are made on the basis of a coarse generalized travel time or cost.

\begin{figure}[h!]
    \centering
    \includegraphics[width=0.9\linewidth]{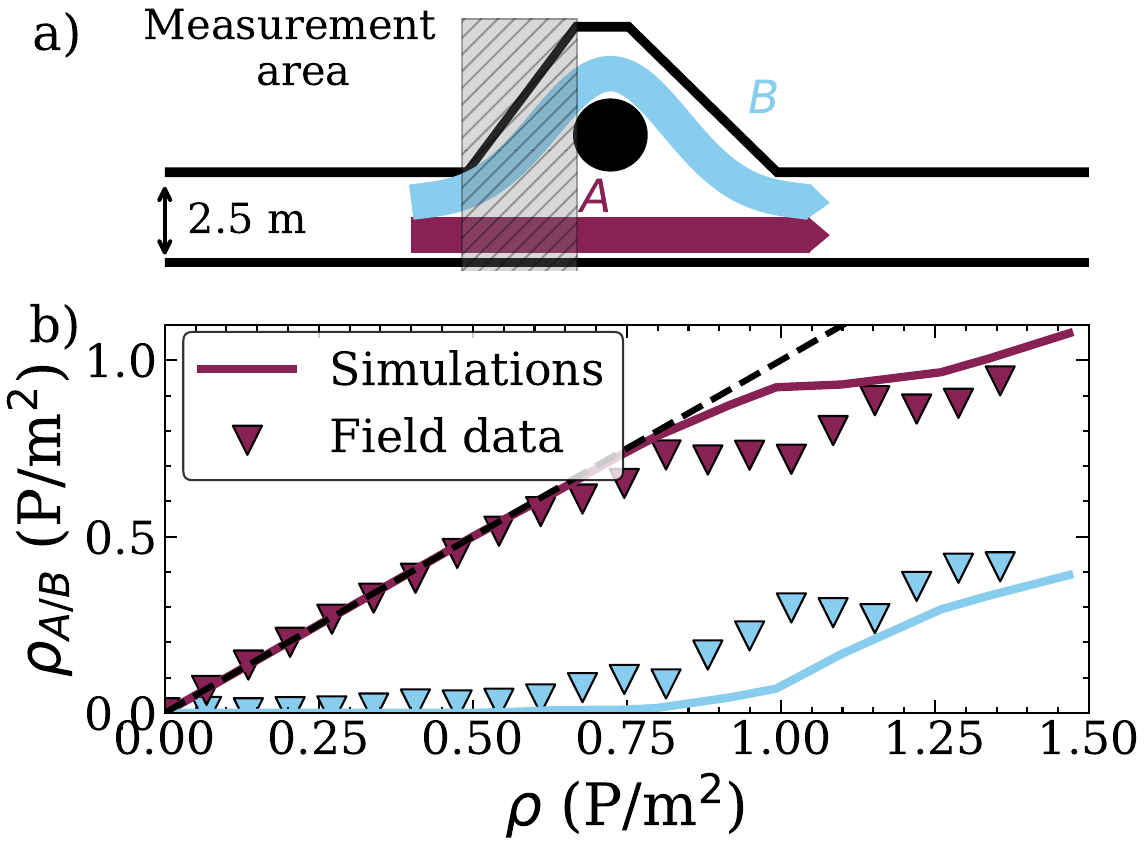}
    \caption{Route choice depending on the density. {\bf (a)} Sketch of the geometry studied on the field by Gabbana et al.~\cite{Gabbana2022Sep} and reproduced numerically here (with periodicity along the horizontal axis). Pedestrians walk to the right and may choose a shorter, but possibly more congested route (\emph{A}) or a longer one (\emph{B}); the two are separated by a support pillar (\emph{black disk}). {\bf (b)} Densities of pedestrians selecting routes \emph{A} and \emph {B}, as a function of the total pedestrian density in the measurement area. The dashed line indicates $\rho_{A/B} = \rho$.}
    \label{fig:largescale}
\end{figure}

Here, we show that the seamless articulation between the operational and tactical levels in the proposed framework efficiently solves modeling issues in concrete situations. For that purpose, we turn to the field study conducted by \citeauthor{Gabbana2022Sep}~\cite{Gabbana2022Sep} at the GLOW light festival in Eindhoven, in the Netherlands. As illustrated in Fig.~\ref{fig:largescale}a, a unidirectional pedestrian flow is confronted with an asymmetric route choice dilemma, either going straight (route $A$) or making a detour around a large support pillar (route $B$). While virtually all pedestrians choose route $A$ at low enough density, a growing fraction select the longer route $B$, as the crowd gets denser and congestion emerges. This choice is anchored in tactical reasoning that extends over at least $\sim 10$ seconds and meters, balancing travel distance against the expected local density.

Figure~\ref{fig:largescale}b proves that, without any specific adjustment, the model reproduces the empirical trends: Below an intermediate density $\rho \approx \SI{0.8}{\mathrm{Ped}\per\meter\squared}$, virtually everybody chooses route $A$, whereas above this threshold more and more pedestrians opt for route $B$, with proportions that are broadly captured by the model. This shows that our mean-field approach is able to fill the gap between the operational and the tactical level.
The mean-field, `tactical' approach implemented beyond $\TantBare$ (Sec.~\ref{sec:meanfield}) was crucial to reproducing these observations. When pedestrians are still $\SI{5}{\meter}$-$\SI{10}{\meter}$ upstream of the pillar, the last atom of their polymer chain, at $t' = t+\TantBare$, reaches the bifurcation point and is guided towards either of the paths by the gradient of the terminal cost, which relies on a continuous, mean-field description of upcoming densities. Thus, routes are chosen in light of the anticipated congestion along them, rather than on the sole basis of local interactions.

\subsection{Low-speed limit of the cost and halting}
So far, the model ingredients have been kept minimal to show that a fairly generic implementation of the framework is already practically efficient. However, a further asset of the general formalism is that the cost function can readily incorporate additional features in order to account for specific, possibly complex behaviors. Let us first focus on halts along walks. It so happens that pedestrians usually prefer stopping for a few seconds before walking again, rather than walking very slowly (`shuffling their feet').

As a matter of fact, this phenomenon naturally emerges from the model, as soon as one pays attention to the constant in the bio-mechanical cost of Eq.~\ref{eq:biomech_cost}. Indeed, there is actually an extra energy expenditure for setting to walk, even at vanishing speed \cite{ludlow2016energy}. Numerically, handling a discontinuity in the cost at $v=0$ is inconvenient, so we smooth it using a polynomial expression below $v^*=\SI{0.5}{\meter\per\second}$ (see Fig.~\ref{fig:halting}a):

\begin{figure}[h!]
    \centering
    \includegraphics[width=0.9\linewidth]{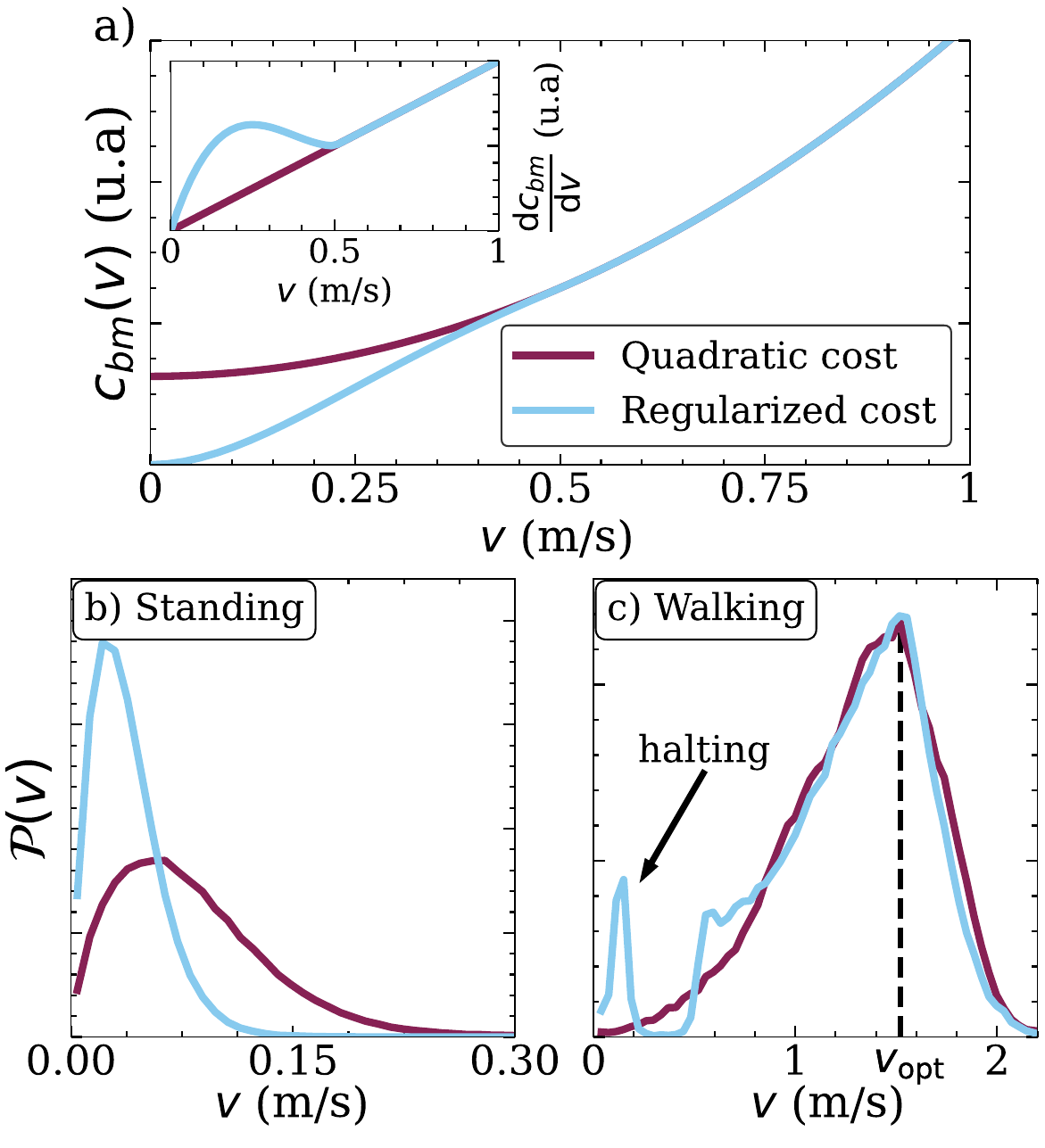}
    \caption{ {\bf (a)} Bio-mechanical cost $c_{bm}(v)$ without drop at $v=0$ (`Quadratic cost') or with a smoothed discontinuity at $v=0$ (`Regularized cost'), as defined in Eq.~\ref{eq:fullbmcost}. \emph{Inset:}  derivative of the cost (used in the cost gradient).
{\bf (b–c)} Simulation of walking agents crossing a standing crowd located in a $17\,\mathrm{m} \times 3\,\mathrm{m}$ rectangle. Probability distributions of the speeds of  {\bf (b)}  standing agents, {\bf (c)} crossing agents.}
    \label{fig:halting}
\end{figure}

\begin{align}
\label{eq:fullbmcost}
    c_{bm}(v) =
    \begin{cases}
        2\alpha{v^*}^2\Big(\Big(\dfrac{v}{v^*}\Big)^4-3\Big(\dfrac{v}{v^*}\Big)^3+3\Big(\dfrac{v}{v^*}\Big)^2\Big), & \text{if } v < v^* \\
        \alpha (v^2+{v^*}^2), & \text{otherwise.}
    \end{cases}
\end{align}

The ensuing steep derivative at low speed (inset of Fig.~\ref{fig:halting}a) has an effect similar to a discontinuity.

Simulating a stream of pedestrians walking through a group of standing pedestrians in a rectangular area,
we find a distribution of speeds that markedly differs depending on whether drop of the bio-mechanical cost at $v=0$ is discarded (Eq.~\ref{eq:biomech_cost}) or smoothly integrated (Eq.~\ref{eq:fullbmcost}). In the former case, standing agents (Fig.~\ref{fig:halting}b) slowly drift due to the social interactions, whereas they stay almost still in the latter case.
Walking agents (Fig.~\ref{fig:halting}c) are also affected: agents that yield to let another one pass first may exhibit arbitrary low speeds in the range $v = \SI{0}{\meter\per\second}$ to $v = v^{\star}$ if the cost drop at standstill is not implemented,
whereas they halt when it is. Accordingly, the `regularization' of the cost makes it more favorable to stop to let another pedestrian pass than to shuffle one's feet all along, and (for agents at rest) to not move unless the perturbation is large enough.
 
\subsection{Applied example of a metro exit}
\label{sec:metroexit}
\begin{figure*}
    \centering
    \includegraphics[width=0.9\linewidth]{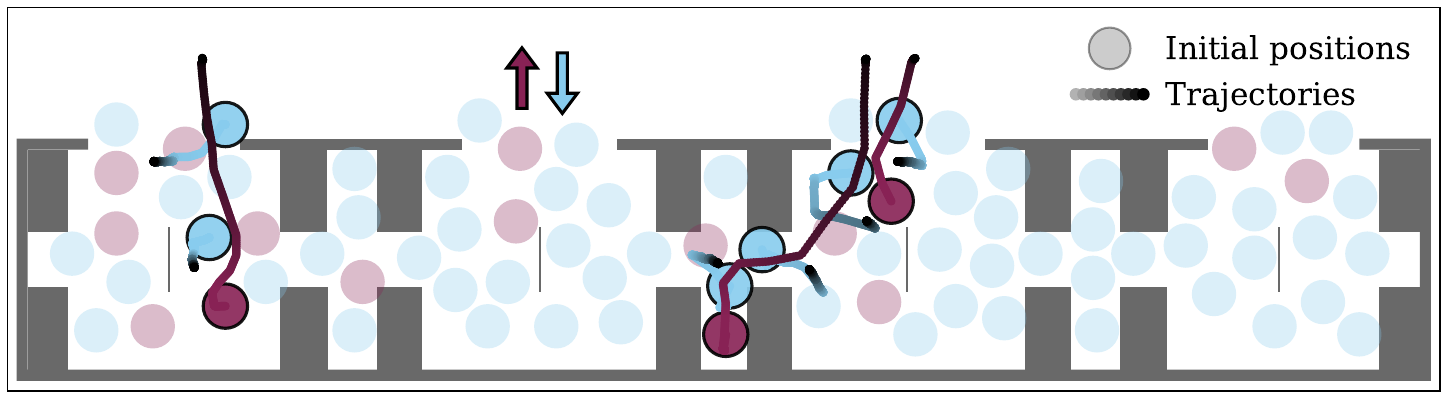}
    \caption{Simulations of passengers alighting from a realistic crowded subway train, whose walls and seats are shown in gray. 80 agents are initially randomly distributed in the train and 16 of them (\emph{in red}) intend to alight; the others (\emph{in blue}) want to stay onboard. The trajectories of a selection of pedestrians are shown to illustrate the avoidance behavior of the agents in blue and the paths followed by agents in red. For red agents, the coefficient $\kappa$ in the final cost increases linearly with time, accounting for the growing urgency to leave the train before the doors close.}
    \label{fig:metroexit}
\end{figure*}

Finally, we consider the integration of impatience, or time-dependent motivation \cite{usten2026well}, to render a complex scenario of direct relevance for applications of crowd dynamics to the design of transport infrastructures: alighting and boarding a subway train. A real (two-dimensional) train geometry is reproduced in Fig.~\ref{fig:metroexit} and a crowd of 80 passengers is randomly placed inside it, as though it were morning rush hour. When the doors open, at $t=0$, 16 passengers will strive to alight while the others want to stay onboard. To account for the mounting pressure to alight or to be onboard before the doors close, at $t=T_\mathrm{close}$, we make the prefactor $\kappa$ in  the target cost (Eq.~\ref{eq:cost_for_peds}) time-dependent:
\begin{align}
    \kappa(t) = \kappa_0 \cdot \Big(1 + 2t/T_\mathrm{close}\Big)^2.
    \label{eq:kappa_t}
\end{align}
In the absence of interactions, this leads to a time-dependent optimal speed $v_\mathrm{opt} = v_0(1+2t/T_\mathrm{close})$; passengers may $\textit{push}$ three times stronger to elbow their way to the exit as $t$ reaches $T_\mathrm{close}$.

Interesting features can be observed in the dynamics shown in the Supplemental video and sketched in Fig.~\ref{fig:metroexit}. Overall, the dynamic patterns in this complex scenario are quite realistic, as far as one can judge with the naked eye. In detail, we observe that some passengers far from the doors, such as the purple one between two seats, eventually become impatient enough, i.e., reach a high enough value of $\kappa$, to manage to make their way to the exit through a large and dense passenger crowd.

Equally interesting is the seemingly cooperative motion of non-alighting passengers who move transversely to give leeway to alighting agents, or may even choose to temporarily step out of the train. These behaviors are obtained without explicit incentives for cooperation, nor any specific psychological factor apart from Eq.~\ref{eq:kappa_t}. Unlike in other models, the agents are not shoved aside by others pushing them, but act in light of what they anticipate. Note that transverse motion to let an intruder through had already been captured using a mean-field game approach \cite{bonnemain2023pedestrians}, but the finer granularity of our agent-based model represents a leap forward in the simulation of complex scenarios.

% AN: peut-être prendre également un exemple où est introduit un petits puits intermédiaire (moins profond que le puits final) pour décrire une possible activité intermédiaire. Le lèche-vitrine pourrait être un exemple, qui précédemment requérait force détours pour être simulé (nouvel objectif, probabilité de changer d'objectif, etc.). À voir si ça marche facilement ou si c'est compliqué.
%Borgers, A. W., Smeets, I. M. E., Kemperman, A. D. A. M., & Timmermans, H. J. P. (2006). Simulation of micro pedestrian behaviour in shopping streets. Progress in design & decision support systems in architecture and urban planning, 101-116.
\section{Conclusion}

Starting from fundamental considerations, we have introduced a delineation between reactive agents and anticipatory ones, whose present-time dynamics depend on the prospective system state as they anticipate it thanks e.g. to an internal model. These dynamics can be expressed in terms of a `cost' function; we expose how this cost, rather than being postulated \emph{a priori} under of a rational hypothesis, can be constructed on the basis of observations. In this framework, the dynamics of an anticipatory agent in $d$ dimensions can be mapped onto the dynamics of a (non-anticipatory) chain in $d+1$ dimensions; the seemingly complex problem of concurrent path planning by multiple agents thus reduces to the search for \emph{satisficing} chain configurations in an augmented space of dimension $d+1$. Uncertainty about the future state of the system is mirrored by fluctuations acting transversely on the chain, more precisely, by subjecting the successive positions that form the chain (once discretized into a polymer-like object) to non-uniform temperatures.

Building on this polymer analogy opened the door  to an analytical characterization of the fluctuations and temporal dynamics of these objects, notably in terms of a dimensionless Péclet number. In the very short term, i.e., up to an anticipation horizon, trajectories are localized enough for inter-agent interactions to be resolved individually. The blurry future, beyond the anticipation horizon, is handled in a mean-field way. 

We applied the resulting unified framework to pedestrian dynamics. Thanks to the consistent bridge between short-term and long-term anticipation, the operational and tactical levels have been integrated seamlessly in an agent-based model which naturally accounts for tactical processes such as route choice while maintaining the granularity that is necessary to resolve complex scenarios. 
Even with a simple, generic expression of the cost, the model was shown to perform well and reproduce experimental results in several scenarios that were particularly challenging to existing models, such as cluttered environments; agents are able to sacrifice their very short-term interests, keeping an eye on their final goal. Interestingly, the effective anticipation horizon is not fixed, but gets shorter at higher density, reflecting the lower predictability of the system. 

The complex trajectories generated by the model do not stem from intricate specifications of the cost or profuse rules, but emerge from the collective anticipatory dynamics. Nonetheless, specific features can readily be incorporated into the cost to describe particular situations. For example, including proximity costs to maintain the cohesion of social groups would be a valuable future extension. The mean-field approach, beyond the anticipation horizon, also suffers from some limitations, which could be remedied, to begin with, by enforcing advection of the density field.

From a fundamental perspective,  being able to mirror `psychological processes' such as the perception of neighbors or anticipation with well-known physical processes in space-time instills the hope that the machinery of Physics can be deployed to study the dynamical implications of these processes. Our findings open up new perspectives on anticipatory systems that could benefit diverse fields of research, from Active Matter Physics to Animal Behavior and Robotics. If the profuse past studies on their counterparts, namely, dynamical systems with memory, are anything to go by, vast swaths of future research territories may open for anticipatory systems.

\begin{acknowledgments}

We acknowledge financial support from Agence Nationale de la Recherche through project MUTATIS (ANR-24-CE22-0918).
AN thanks Jakob CORDES for his early input into the project.

\end{acknowledgments}

\FloatBarrier

\onecolumngrid
\newpage
\appendix
\renewcommand{\thefigure}{A\arabic{figure}}
\setcounter{figure}{0}

\section{Calculations of the mean-squared displacement of a polymer chain in a temperature gradient}
\label{app:calc_RMSD}

\subsection{Projection of the atom displacements on the dynamic eigenmodes}
The dynamics of the $N+1$ `atoms' forming the polymer, pinned at $\boldsymbol{r}_0=\boldsymbol{0}$ at one end and pulled by a force at the other end, are governed by the 
$N$ linear stochastic differential equations of Eq.~\ref{overdamped_langevin} of the main text.
We look for coefficients $\alpha_p^{(n)}$ which diagonalize the matrix $M_{nm}$ representing the elastic part of these equations:
\begin{equation}
\begin{cases}
  \alpha_{p}^{(n+1)} + \alpha_{p}^{(n-1)} - 2\alpha_{p}^{n} = -\widetilde\lambda_p\alpha_{p}^{(n)}  \\
  \alpha_{p}^{(N-1)} - \alpha_{p}^{(N)} = -\widetilde\lambda_p\alpha_{p}^{(N)},
\end{cases}
\end{equation}
where we have set $\alpha_{p}^{(0)}=0$ for convenience and defined $\widetilde\lambda_p=\frac{dt}{2\alpha}\lambda_p$ (we will drop the tildes in the following). We surmise that $\alpha_p^{(n)} \propto \sin{(k_p n + \phi_p)}$, with $\phi_p=0$ because of the fixed boundary condition. This leads to:
\begin{subequations}
\begin{align}
  \sin{\Big(k_p(n+1)\Big)} + \sin{\Big(k_p(n-1)\Big)} - 2\sin{\Big(k_pn\Big)} &= -\lambda_p\sin{\Big(k_pn\Big)}\\
  \sin{\Big(k_p(N-1)\Big)} - \sin{\Big(k_pN\Big)} &= -\lambda_p\sin{\Big(k_pN\Big)}.
  \label{eq:sinsin}
\end{align}
\end{subequations}
By factorizing the sum of the two sines in the first equation and then dividing by $\sin{(k_pn)}$, we get:

\begin{equation}
\lambda_p = 2\cdot\Big(1 - \cos{k_p}\Big)= 4\sin^2\Big(\frac{k_p}{2}\Big).\end{equation}

Rewriting Eq.~\ref{eq:sinsin} as
\begin{align}   
   -2\cos{\Big(k_p(N-\frac{1}{2})\Big)}\sin{\Big(\frac{k_p}{2}\Big)} &= -4\sin^2{\Big(\frac{k_p}{2}\Big)}\sin{\Big(k_p N\Big)} \nonumber \\
   \cos{\Big(k_p(N-\frac{1}{2})\Big)} &= 2\sin{\Big(\frac{k_p}{2}\Big)}\sin{\Big(k_p N\Big)} \nonumber \\
    \cos{\Big(k_p(N-\frac{1}{2})\Big)} &= \cos{\Big(k_p N-\frac{k_p}{2}\Big)}-\cos{\Big(k_p N+\frac{k_p}{2}\Big)} \nonumber \\
    \cos{\Big(k_p \cdot\frac{2N+1}{2}\Big)}&=0,
\end{align}
we arrive at:
\begin{equation}
    k_p = \frac{2p+1}{2N+1}\pi; ~~~ \alpha_p^{(n)} \propto \sin{\Big(\frac{2p+1}{2N+1}n\pi\Big)}.
\end{equation}
Finally, we choose the following normalization
\begin{equation}
    \alpha_p^{(n)} = \frac{2\ \sin{\Big(\frac{2p+1}{2N+1}n\pi\Big)}}{\sqrt{2N+1}},
\end{equation}
so that $\sum_{n=0}^{N}\alpha_p^{(n)}\alpha_q^{(n)}= \delta_{pq}$, as we prove in the next subsection.

\subsection{Calculation of the normalization constant for the $\alpha_p^{(n)}$}

With the foregoing normalization, we verify that, for all non-negative integers $p,\,q$, 
\begin{align}
    \sum_{n=0}^{N}\alpha_p^{(n)}\alpha_q^{(n)} &= \sum_{n=0}^{N}  \frac{4}{2N+1} \cdot\sin{\Big(\frac{2p+1}{2N+1}n\pi\Big)}\cdot\sin{\Big(\frac{2q+1}{2N+1}n\pi\Big)} \label{eq:normalization} \\
    &= \frac{4}{2N+1} \cdot  \frac{-1}{4}\sum_{n=0}^{N} \Big\{e^{{\bf i}sn} + e^{-{\bf i}sn}  -e^{{\bf i}dn} - e^{-{\bf i}dn} \Big\} \nonumber \\
     &= \frac{-1}{2N+1}\Big\{\sum_{n=-N}^{N} \Big[e^{{\bf i}s}\Big]^n - \sum_{n=-N}^{N} \Big[e^{{\bf i}d}\Big]^n \Big\}  \nonumber\\
     &= \frac{1}{2N+1} (2N+1) \delta_{pq} \nonumber  \\
     &=\delta_{pq}, \nonumber 
    \end{align}
where we have introduced the shorthands $s \hat{=} \frac{2(p+q+1)}{2N+1}\pi$ and $d \hat{=} \frac{2(p-q)}{2N+1}\pi$.

\subsection{Correlation of the noise modes with quadratic temperature}
%\ref{}
Let us calculate the correlation between two modes $p$ and $q$ of the noise term in our Langevin equation Eq.~18:
\label{correlation_quadratic}
    \begin{align}
    \langle \tilde{\eta}_p(\tau)\tilde{\eta}_q(\tau')\rangle &= \sum_{n=0}^{N} \sum_{m=0}^{N}\alpha_p^{(n)}\alpha_q^{(m)}\langle\eta_n \eta_m\rangle\\
    &=4\Theta \dd t^2\sum_{n=0}^{N}\sum_{m=0}^{N}\alpha_p^{(n)}\alpha_q^{(m)}\cdot n\cdot m\cdot\delta_{nm}\cdot\delta(\tau-\tau')\\
    &=\Theta \dd t^2\delta(\tau-\tau')\,\sum_{n=0}^{N} 4\,n^2 \alpha_p^{(n)}\alpha_q^{(n)}.
    \end{align}

Now, let us introduce
\begin{equation}
S(s,d)= \frac{-1}{2N+1}\Big\{\sum_{n=-N}^{N} e^{{\bf i}sn} - \sum_{n=-N}^{N} e^{{\bf i}dn} \Big\}
\end{equation}
and notice that,  in the light of Eq.~\ref{eq:normalization}, for $p \neq q$,
 \begin{align}
g_{pq} &\hat{=} \sum_{n=0}^{N} 4\,n^2 \alpha_p^{(n)}\alpha_q^{(n)} \nonumber \\
&= - 4 \frac{\partial^2 S}{\partial s^2} - 4 \frac{\partial^2 S}{\partial d^2} \nonumber \\
&= \frac{4}{2N+1} \cdot \Big\{\frac{\partial^2 }{\partial s^2}\Big[ \frac{\sin( (N+\frac{1}{2})s)}{\sin(\frac{s}{2})} \Big] - \frac{\partial^2 }{\partial d^2} \Big[ \frac{\sin( (N+\frac{1}{2})d)}{\sin(\frac{d}{2})} \Big]  \Big\} \nonumber \\
&= 4 \cdot \Big\{ 0+ \frac{-\cos\Big((N+\frac{1}{2})s\Big)\cos(\frac{s}{2})}{2\sin^2(\frac{s}{2})} + \frac{\cos\Big((N+\frac{1}{2})d\Big)\cos(\frac{d}{2})}{2\sin^2(\frac{d}{2})}\Big\} \nonumber \\
&= 2 \cdot (-1)^{p+q} \cdot  \Big[ \frac{\cos(\frac{s}{2})}{\sin^2(\frac{s}{2})} + \frac{\cos(\frac{d}{2})}{ \sin^2(\frac{d}{2})}\Big],
\end{align}
where we have used the fact that, at the evaluation point, $\sin( (N+\frac{1}{2})s)=0$ and $\sin((N+\frac{1}{2})d)=0$. In the limit $p,\,q\ll N$,
\begin{equation}
   g_{pq} \approx (-1)^{p+q} \cdot  \frac{2}{\pi^2} \cdot (2N+1)^2 \cdot \Big[\Big(\frac{1}{p+q+1}\Big)^2 + \Big(\frac{1}{p-q}\Big)^2\Big]
\end{equation}

For $p=q<N$, as $d=0$, the second term in $S(s,d)$ cannot be expressed in the same way. Instead, we have
\begin{align}
g_{pp} &\hat{=} 4\sum_{n=0}^{N} n^2 \alpha_p^{(n)}\alpha_p^{(n)} \nonumber \\
&= - 4\frac{\partial^2 S}{\partial s^2} + \frac{4}{2N+1}\sum_{n=0}^{N}2n^2\\
&=  2 \cdot \frac{\cos(\frac{s}{2})}{\sin^2(\frac{s}{2})}  +\frac{4}{3}\cdot N\cdot(N+1). 
\end{align}
In the limit $p,\,q\ll N$,
\begin{equation}
    g_{pp} \approx  \frac{2}{\pi^2} \cdot\Big(\frac{2N+1}{2p+1}\Big)^2 +\frac{4\cdot N\cdot(N+1)}{3}.
\end{equation}

Finally, the case $p=q=N$ is straightforward: $g_{NN}=0$.

\subsection{Mean-square displacements}

We can now integrate the Ornstein-Ulhenbeck process (Eq.~\ref{eq:isolated_polymer_transformed}) to find the correlation between modes $p$ and $q$ (in $p$-space):

    \begin{align}
        \langle(u_p(\tau) - u_p(0))((u_q(\tau) - u_q(0))\rangle &= \int_0^\tau\int_0^\tau\langle \tilde{\eta}_p(\tau')\tilde{\eta}_q(\tau'')\rangle e^{-\frac{\tau-\tau'}{\tau_p}}e^{-\frac{\tau-\tau''}{\tau_q}}\dd \tau' \dd \tau'' \nonumber \\
        &= g_{pq}\Theta \dd t^2\int_0^\tau e^{-(\tau - \tau')(\frac{1}{\tau_p} + \frac{1}{\tau_q})}\dd \tau' \nonumber\\
        &= g_{pq}\Theta \dd t^2\int_0^\tau e^{-(\tau - \tau')(\kappa_p + \kappa_q)}\dd \tau' \nonumber\\
        &= \frac{g_{pq}}{\kappa_p+\kappa_q}\Theta\dd t^2\Big[1-e^{-\tau (\kappa_p + \kappa_q)}\Big].
    \end{align}
This enables us to calculate the mean-squared displacement of the $n$-th atom in real space:
    \begin{align}
        \Big\langle(r_n(\tau) - r_n(0))^2\Big\rangle &= \sum_{p=0}^N\sum_{q=0}^N\alpha_p^{(n)}\alpha_q^{(n)}\Big\langle\left(u_p(\tau) - u_p(0)\right)\left((u_q(\tau) - u_q(0)\right)\Big\rangle \nonumber\\
        &= \Theta \dd t^2\sum_{p=0}^N\sum_{q=0}^N \alpha_p^{(n)}\alpha_q^{(n)}\frac{g_{pq}}{\kappa_p+\kappa_q}\Big(1-e^{-\tau (\kappa_p + \kappa_q)}\Big).
    \end{align}
Thus, after the initial transient, when $\tau\to \infty$, the mean-square displacement tends to 
\begin{align}
   \Delta r_n^2 &\approx \Theta \dd t^2 (\alpha_0^{(n)})^2g_{00}\frac{\tau_0}{2}  \nonumber \\
        &\approx \left(\frac{4}{\pi^2} + \frac{2}{3} \right)\, \Theta \, \dd t^2 n^2\frac{N\dd t}{\alpha},
\end{align}
where the last approximation holds for $n\ll N$.

\section{Complete set of simulations corresponding to the static crowd crossing experiments}
\label{appendix:wang}
\begin{figure*}[h!]
    \centering
    \includegraphics[width=1\linewidth]{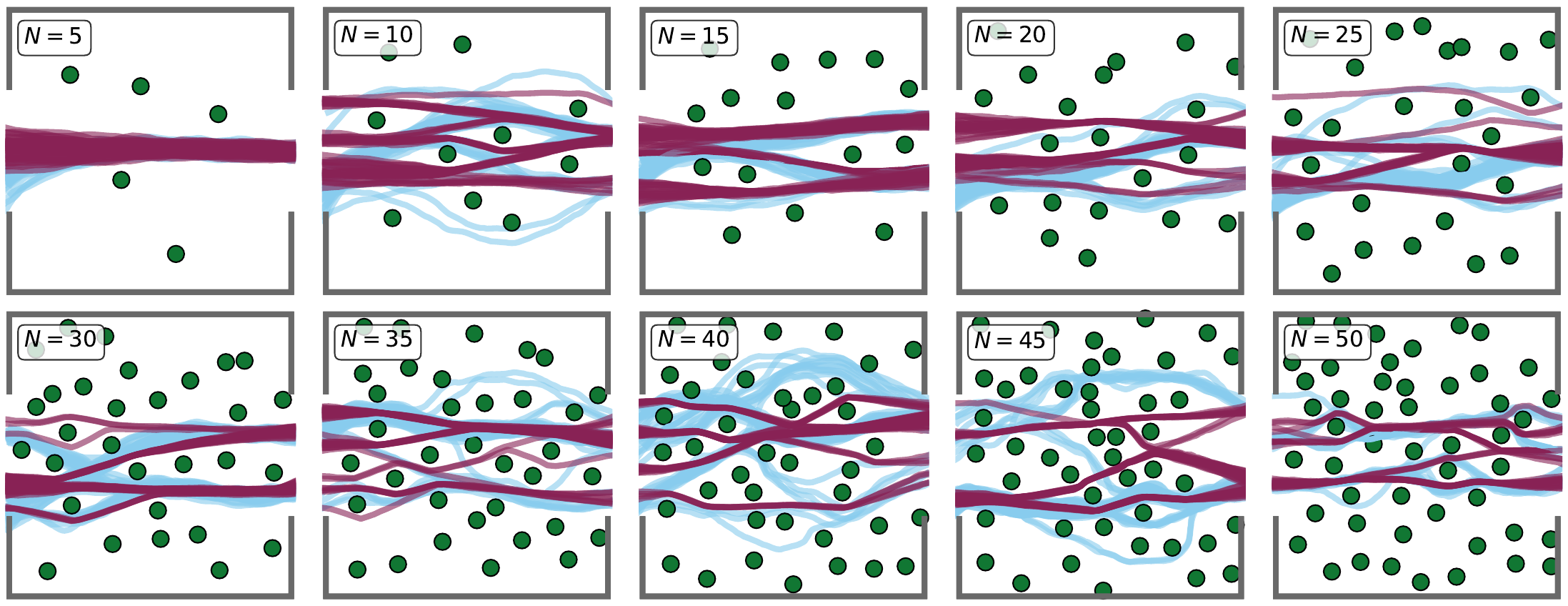}
    \caption{Experimental trajectories measured  by Wang et al. \cite{wang2023exploring} (\emph{in blue}) and their numerical counterparts (\emph{in red}) at various densities. Refer to Fig.~\ref{fig:crossing} for the rest of the caption. }
    \label{fig:wang_full_traj}
\end{figure*}

\begin{figure*}[h!]
    \centering
    \includegraphics[width=\linewidth]{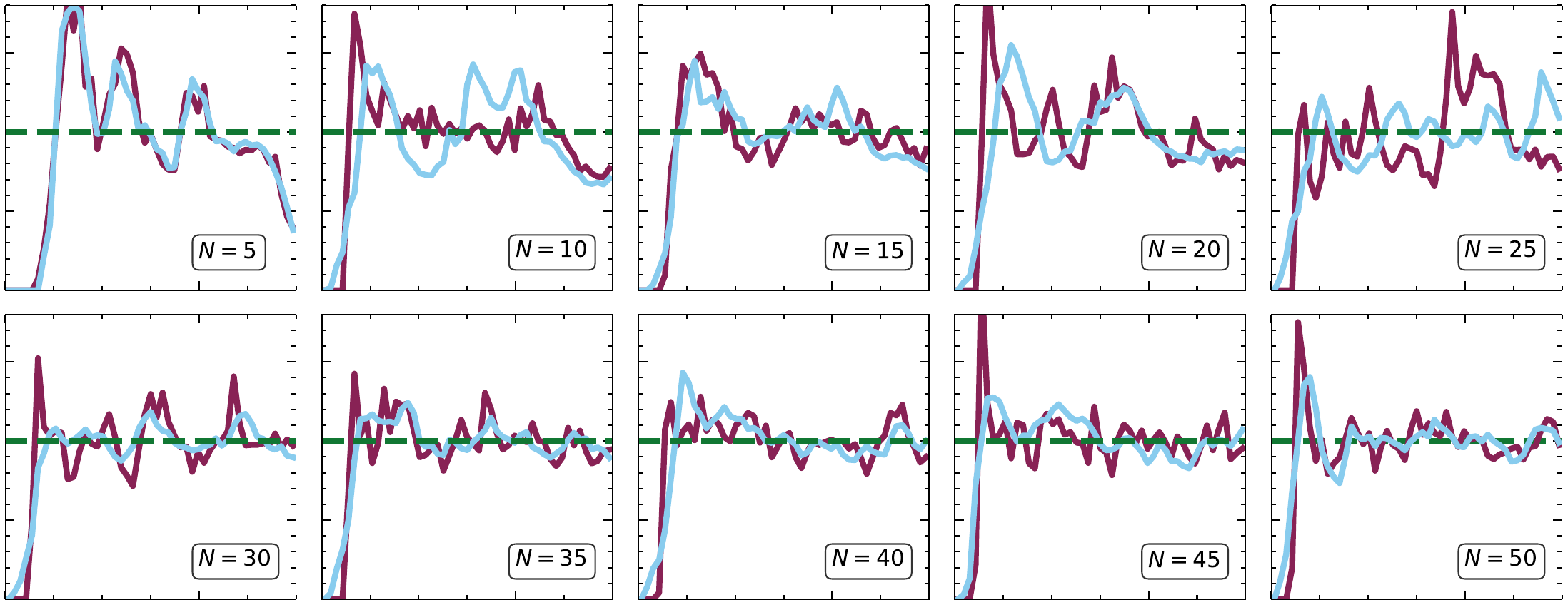}
    \caption{Pair distribution functions (pdf) $g(r)$ between the crossing pedestrian and the static ones, computed from experimental data (\emph{in blue}) and numerical simulations (\emph{in red}), at various densities. Refer to Fig.~\ref{fig:crossing} for the rest of the caption.}
    \label{fig:wang_full_proba}
\end{figure*}
\end{document}